\documentclass{aastex701}
\usepackage{amsmath}
\usepackage{mathrsfs}
\usepackage{graphicx}
\usepackage{subcaption}

\received{...}
\revised{...}
\accepted{...}
\published{...}

\begin{document}
\title{GW230729\_082317 and GW231118\_005626: A Candidate of Hierarchical Triple Mergers}
\author{Bo Gao}
\email{bogao@pmo.ac.cn}
\affiliation{Key Laboratory of Dark Matter and Space Astronomy, Purple Mountain Observatory, Chinese Academy of Sciences, Nanjing 210033, China}
\author[orcid=0000-0001-9120-7733, gname="Shao-Peng", sname="Tang"]{Shao-Peng Tang}
\affiliation{Key Laboratory of Dark Matter and Space Astronomy, Purple Mountain Observatory, Chinese Academy of Sciences, Nanjing 210033, China}
\email{tangsp@pmo.ac.cn}
\author[orcid=0000-0002-8966-6911, gname="Yi-Zhong", sname="Fan"]{Yi-Zhong Fan}
\email{yzfan@pmo.ac.cn}
\affiliation{Key Laboratory of Dark Matter and Space Astronomy, Purple Mountain Observatory, Chinese Academy of Sciences, Nanjing 210033, China}
\affiliation{School of Astronomy and Space Science, University of Science and Technology of China, Hefei, Anhui 230026, China}
\author[gname="Qingzhong", sname="Liu"]{Qingzhong Liu}
\email{qzliu@pmo.ac.cn}
\affiliation{Key Laboratory of Dark Matter and Space Astronomy, Purple Mountain Observatory, Chinese Academy of Sciences, Nanjing 210033, China}

\begin{abstract}

Hierarchical triple mergers denote a sequential merger process in which the black hole remnant from an initial binary merger subsequently merges with a third black hole. In this study, we identify a related pair of binary black hole mergers, GW230729\_082317 and GW231118\_005626, which exhibit consistent sky localization and luminosity distances; the latter is identified as a candidate hierarchical merger binary back holes. The Bayes factor in favor of a hierarchical triple merger origin over the non-association hypothesis is $110$. However, after accounting for the priors odds, the corresponding odds ratio decreases to $0.1$, indicating that the current evidence is insufficient to favor the hierarchical triple merger scenario. GW230720\_0823 show no detectable signatures of orbit eccentricity or line-of-sight acceleration. The inferred properties of these two gravitational wave events suggest a dynamical formation channel, such as few-body scattering or gas-assisted mergers, while a primordial triple origin is disfavored. Among the proposed dynamical environments, the accretion disk of an active galactic nucleus provides the most plausible site. We further derive constraints on the local rate of hierarchical triple mergers rate under both the null and candidate scenarios. Under the assumption of no detected hierarchical triple mergers, the most stringent $90\%$ confidence upper limit is $4.2\,{\rm Gpc^{-3}yr^{-1}}$ for $t_{\rm max}=10$ years using the GWTC-4 catalog.

\end{abstract}


\section{Introduction}\label{sec-int}

The first part of the fourth observing run (O4a) of the LIGO/Virgo/KAGRA (LVK) Collaboration was conducted from 24 May 2023 to 16 January 2024. During O4a, 128 significant gravitational wave (GW) candidates were identified, substantially increasing the size of the LVK event catalog \citep{2026ApJ..1004L..22A}. Including previous observing runs, the current catalog contains $218$ GW events as of O4a, the vast majority of which are binary black hole (BBH) mergers, together with two binary neutron star (BNS) mergers and seven neutron star-black hole (NSBH) mergers. An increasing number of detected systems exhibit unusual properties, including black holes with masses in or near the pair-instability mass gap, unusually low masses, or highly asymmetric mass ratios, posing significant challenges to conventional isolated binary evolution scenarios \citep{2020ApJ...896L..44A,2020ApJ...900L..13A,2024ApJ...970L..34A,2025ApJ...993L..25A}. As the GW catalog continues to expand, increasingly rare astrophysical phenomena are expected to become observable, including hierarchical mergers \citep[e.g.,][]{2002MNRAS.330..232C,2021ApJ...915L..35K,2021arXiv211010838W,2022ApJ...941L..39W,2024PhRvL.133e1401L,2026ApJ...999..127L,2025arXiv250923897L,2026arXiv260701121L}, mergers originating from hierarchical triple systems \citep{2020MNRAS.498L..46V}, and gravitationally lensed GW events \citep{2024ApJ...970..191A}. These discoveries provide valuable opportunities to distinguish between competing formation channels \citep{2017CQGra..34cLT01V,2020A&A...635A..97B}, to probe the dynamical evolution of compact objects in dense environments \citep{2021NatAs...5..749G,2026arXiv260320430A}, and to test fundamental BH physics \citep{2025PhRvL.135k1403A, 2026SciBu..71...83T, 2026PhRvL.136d1403A, 2026arXiv260816563T}. In this context, dynamical formation channels are expected to play an increasingly important role in explaining the observed GW population. However, current theoretical predictions remain subject to substantial uncertainties arising from both the treatment of few-body dynamical processes and the poorly constrained properties of their host environments \citep{2025ApJ...991..131Z}, such as the black hole mass function \citep{2020PhRvD.102l3016A}, natal kick and escape velocities \citep{2021MNRAS.505..339M}, and the orbital parameter distributions of hierarchical triples \citep{2022MNRAS.511.1362T,2026MNRAS.545f1938D}. 

Triple systems provide a natural environment for a broad range of dynamical and astrophysical phenomena. Secular Lidov-Kozai oscillations and resonant few-body interactions can excite the orbital eccentricity of binaries, accelerate their GW driven inspiral, and in some cases produce successive mergers. Consequently, hierarchical triples have been proposed as a promising formation channel for compact binary mergers detectable by ground-based GW observatories \citep{2016ApJ...831..187A,2022MNRAS.511.1362T,2025A&A...699A.272V,2026MNRAS.545f1938D}. Motivated by this possibility, considerable effort has been devoted to searching for observational evidence of hierarchical triple merger in the LVK catalogs. Searches for correlated GW events have relied on the consistency of merger remnant properties, sky localization, and merger ordering, and the expected statistical distributions of hierarchically formed black holes \citep{2018arXiv180707062H,2020MNRAS.498L..46V,2021ApJ...907L..48V,2023arXiv230709097G,2025PhRvD.111j3016L}. Several GW event pairs, including GW151012-GW170809, GW190814-GW200210\_092254 and GW190514-GW190521, have been proposed as potential hierarchical triple merger candidates. However, none of these candidates has yet reached sufficient statistical significance to establish hierarchical triple merger origin unambiguously. As the GW catalog continues to grow, correlated pairs of mergers produced by hierarchical triples are expected to become increasingly accessible, enabling direct tests of this formation channel.

We performed a search for hierarchical triple merger candidates using the O4a GW dataset. Here, we report two correlated mergers discovered in 2023, GW230729\_082317 and GW231118\_005626, whose binary black hole mergers may be linked through a hierarchical triple system or through the dynamical recapture of a merger remnant. This hypothesis is consistent with previous work that the primary black hole of GW231118\_005626 has been classified as a second generation merger, whereas GW230729\_082317 is a first generation merger \citep{2025arXiv250923897L,2026SCPMA..6999562W}. The main content of this paper is organized as follows. Section~\ref{sec:obs} details the observational properties of GW230729\_082317 and GW231118\_005626. In Section~\ref{sec:form}, we discuss potential formation scenarios, and in Section~\ref{sec:rate} we derive an upper limit on the hierarchical triple merger rate. Our conclusions are summarized in Section~\ref{sec:concl}.

\section{Information about GW230729\_082317 and GW231118\_005626}\label{sec:obs}

A search for event pairs in GWTC-4 and earlier catalogs was performed by comparing the posterior distributions of extrinsic (sky location, luminosity distance) and intrinsic (mass, spin) parameters. Candidate associations were examined with a Bayesian framework described in previous work \citep{2024ApJ...965...80G}, revealing GW230729\_082317 and GW231118\_005626 as the most strongly correlated pair.

GW230729\_082317 is a binary black hole merger detected by LVK during the O4a observing run. It was observed with a signal-to-noise ratio (SNR) of $9.5$ and a low false alarm rate (FAR) of $3.4 \times 10^{-9}$ Hz, corresponding to approximately one false alarm every $9.35$ years\footnote{\url{https://gracedb.ligo.org/superevents/S230729z/}}, providing high statistical credible in its astrophysical origin. The system consisted of two black holes with median source-frame masses of approximately $12.3\,M_\odot$ and $7.6\,M_\odot$, which merged to form a remnant black hole of $\sim 19.4\,M_\odot$ at a luminosity distance of $\sim 1630$ Mpc. The effective spin is $0.12$. Its exceptionally low FAR and small effective spin make this event a clean observational sample for probing dynamical formation channels of binary black hole. GW231118\_005626 is another binary black hole merger detected during O4a observing run. The system exhibits a pronounced mass asymmetry, with a primary-to-secondary mass ratio of approximately $2:1$, making it potential carrier of dynamical information among the BBH system in the GWTC-4 catalog. The FAR corresponds to approximately one false alarm every $100.04$ years\footnote{\url{https://gracedb.ligo.org/superevents/S231118d/}}, indicating a low probability of a noise origin. The event has a SNR of $10.7$ and an effective spin of $\sim 0.4$, suggesting a moderately spinning system. Its low FAR and unusual mass ratio make it a valuable system for investigating BBH formation channels, particularly those involving dynamical assembly.

\subsection{Posterior Distributions of Key Parameters about the Double Mergers}

\begin{figure}[htbp]
    \centering
    \includegraphics[width=0.5\linewidth]{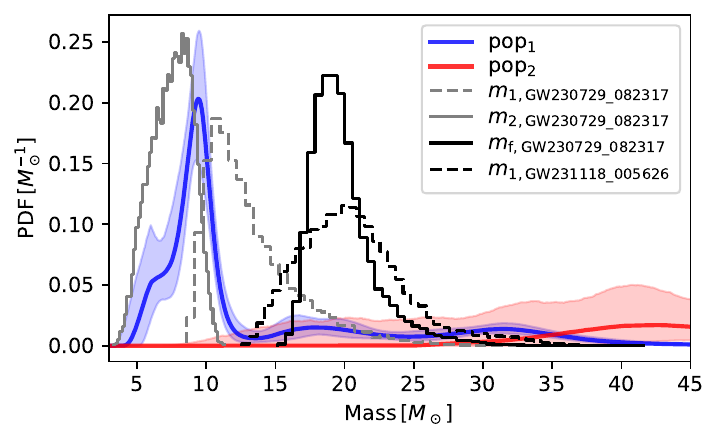}
    \caption{Primary and secondary mass distributions for GW230729\_082317, together with the primary mass distribution for GW231118\_005626. The blue and red curves denote the mass distributions of subpopulation 1 and 2, which are commonly considered to be first-generation BHs and the remnants of previous mergers, respectively. The shaded areas mark the $90\%$ credible regions. For ease of comparison, the mass distribution of subpopulation 2 shown in the figure has been scaled by a factor of $20$.}
    \label{fig:mass}
\end{figure}

\begin{figure}[htbp]
    \centering
    \includegraphics[width=0.5\linewidth]{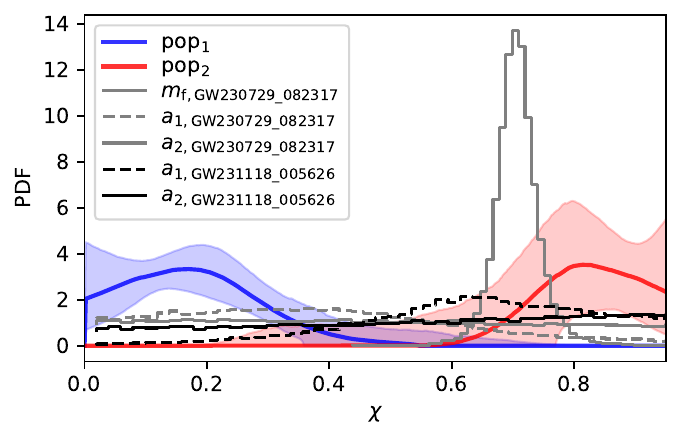}
    \caption{Component Spin distribution of GW230729\_082317 and GW231118\_005626. The blue and red lines represent the spin distribution of the subpopulations, following the same convention as Figure~\ref{fig:mass}.}
    \label{fig:spin}
\end{figure}

\begin{figure}[htbp]
    \centering
    \includegraphics[width=0.5\linewidth]{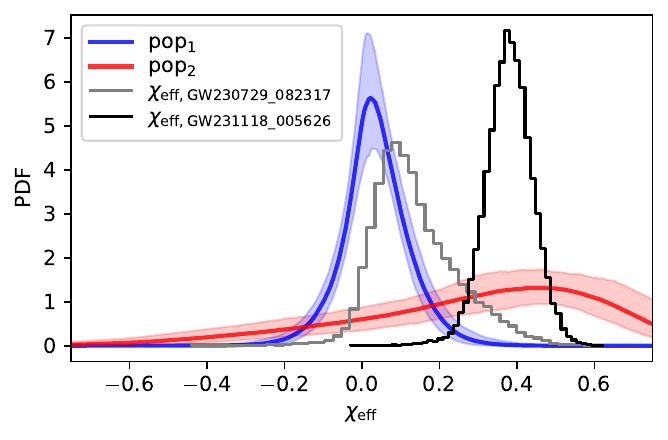}
    \caption{Effective spin distribution of GW230729\_082317 and GW231118\_005626. The blue and red lines represent the effective spin distribution of the subpopulations, following the same convention as Figure~\ref{fig:mass}.}
    \label{fig:spin-eff}
\end{figure}

\begin{figure}[htbp]
    \centering
    \includegraphics[width=0.5\linewidth]{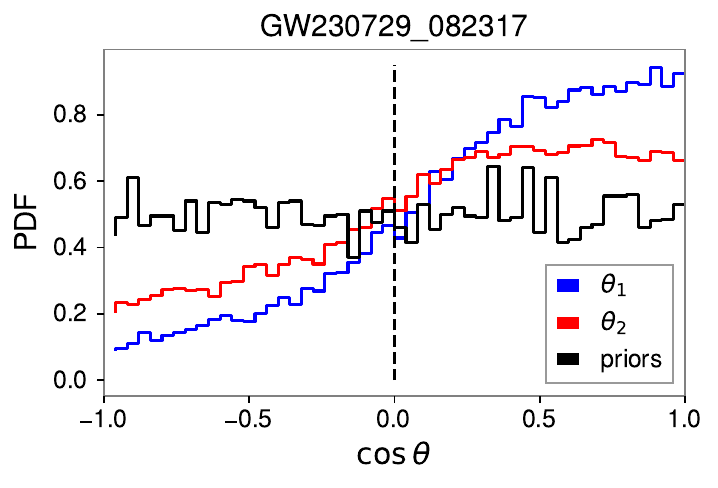}
    \caption{The angle between spin and orbital angular momentum of GW230729\_082317. The black dashed line indicated that spin is perpendicular to orbital angular momentum.}
    \label{fig:spin-inner}
\end{figure}

The component masses of binary systems, directly inferred from GW observations, constitute one of the most informative observables for GW astronomy. They provide essential inference on fundamental physics and astrophysical formation scenarios, while simultaneously offering valuable insights into the origin and evolution of compact binaries. To illustrate the mass relation between the two events, we compare the remnant mass distribution of GW230729\_082317 with the primary mass distribution of GW231118\_005626 in Figure~\ref{fig:mass}. The data shown in Figure~\ref{fig:mass} are taken from the posterior samples released by  the Gravitational Wave Open Science Center (GWOSC)\footnote{\url{https://zenodo.org/records/20275769}} \citep{2026ApJ..1004L..22A,2026arXiv260527225T}. The remnant mass of GW230729\_082317 agrees well with the primary component mass of GW231118\_005626. In addition, the spin distributions for these two events are shown in Figure~\ref{fig:spin}. Owing to the limited sensitivity of current detectors and the spin configurations of the merging black holes, the spin constraints remain weak. Despite the limited precision of the spin magnitude measurements, the peaks of the remnant spin posterior for GW230729\_082317 remains consistent with that of the primary component of GW231118\_005626. Effective spin is a quantity that can be measured with relatively high precision using current GW detectors and is routinely combined with mass to probe black hole formation channels. The effective spin is defined as 
\begin{eqnarray}
    \chi_{\rm eff} \equiv \frac{m_1\boldsymbol{\chi_1}+m_2\boldsymbol{\chi_2}}{m_1+m_2} \cdot \hat{\boldsymbol{L}} \mbox{,}
\end{eqnarray}
where $m_{1,2}$ are the components masses, $\boldsymbol{\chi_{1,2}} = c \boldsymbol{S_{1,2}}/(Gm_{1,2}^2)$ are the dimensionless spins magnitudes, and $\hat{\boldsymbol{L}}$ is the direction of angular momentum. Figure~\ref{fig:spin-eff} displays the effective spin distributions for the two GW events. Across Figure~\ref{fig:mass}-\ref{fig:spin-eff}, the blue and red curves correspond to the mass, spin, and effective spin distributions of the first and second generation mergers, respectively. The masses and spins of the two populations are adopted from \citet{2025arXiv250923897L} based on GWTC-4. The effective spin distributions are derived from the mass, spin, and tilt angle distributions, which is consistent with that inferred from GWTC-3 \citep{2025ApJ...987...65L}. We find that the effective spins of the two events are statistically consistent with the characteristic distributions expected for first and second generation black hole populations. As shown in Figure~\ref{fig:spin-inner}, although the posterior for the spin-orbit inclination is rather broad, the spin of the GW230729\_082317 binary shows a preference for alignment with the orbital angular momentum. Such a preference favors a formation scenario in an AGN disk, where gas torques can efficiently align the spins of the black holes. In contrast, tertiary-induced mergers predict a distinct spin evolution governed by two main processes. First, differential spin precession randomizes the relative orientations of the two black hole spins and the orbital angular momentum, naturally producing a $\chi_{\rm eff}$ distribution peaked near zero. Second, spin-orbit coupling drives the spin-orbit misalignment angles toward $\pi/2$ as the binary hardens under GW emission. The latter effect has been highlighted in previous studies, which find that for nearly equal-mass binaries, the distribution of $\theta_{s,l}$ is peaked around $90^\circ$ \citep{2018MNRAS.480L..58A,2018ApJ...863...68L,2019ApJ...881...41L}. 


\begin{figure}[htbp]
    \gridline{
    \fig{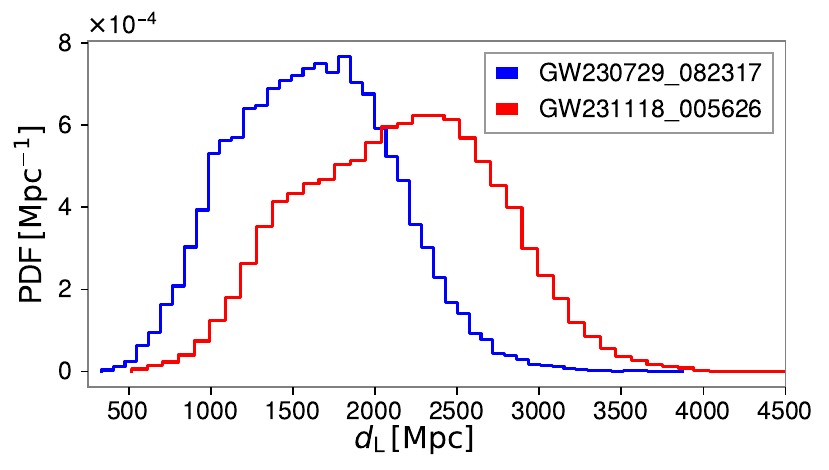}{0.45\textwidth}{(a): Luminosity distance $d_{\rm L}{\rm (Mpc)}$}
    \fig{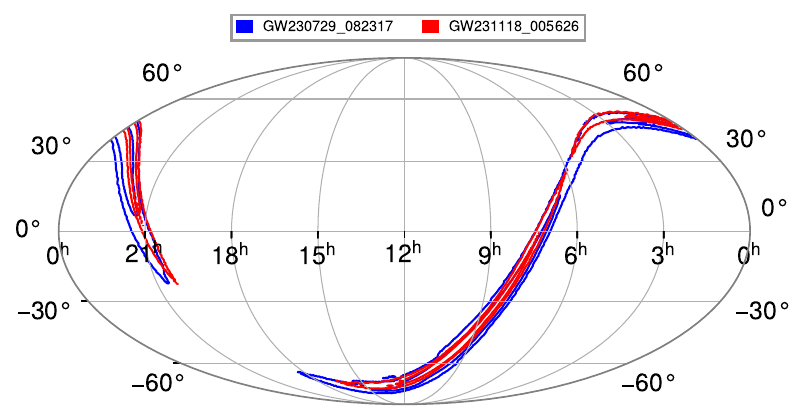}{0.49\textwidth}{(b): Sky localization shown in ICRS equatorial coordinate (RA, Dec, J2000)}}

    \caption{The posterior distributions of extra parameters for GW230729\_082317 and GW231118\_005626 are presented in this figure.}
    \label{fig:sky}
\end{figure}

Motivated by the tentative association suggested by the mass and spin properties, we proceeded to compare the extrinsic parameters of the two events. Figure~\ref{fig:sky} demonstrates that the sky localizations and luminosity distances overlap within their respective credible regions, with a particularly good agreement in the sky positions. This consistency across intrinsic and extrinsic parameters provides strong, multifaceted support for a common origin in hierarchical triple merger.


\begin{figure}[htbp]
    \centering
    \includegraphics[width=0.5\linewidth]{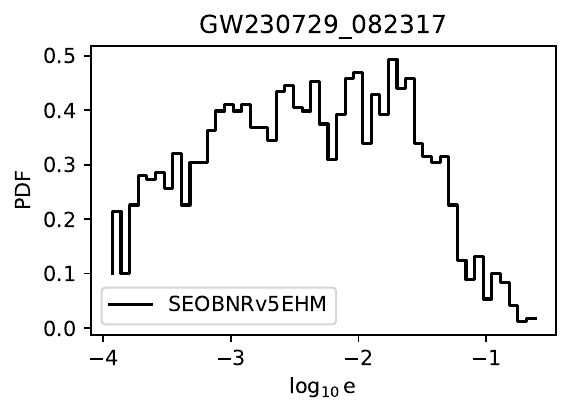}
    \caption{Inferred eccentricity of GW230729\_082317 at $20$ Hz using SEOBNRv5EHM, with the eccentricity sampled from a log-uniform prior over $[10^{-4}, 0.25]$ and displayed on a logarithmic scale.}
    \label{fig:ecc}
\end{figure}

\begin{figure}[htbp]
    \centering
    \includegraphics[width=0.5\linewidth]{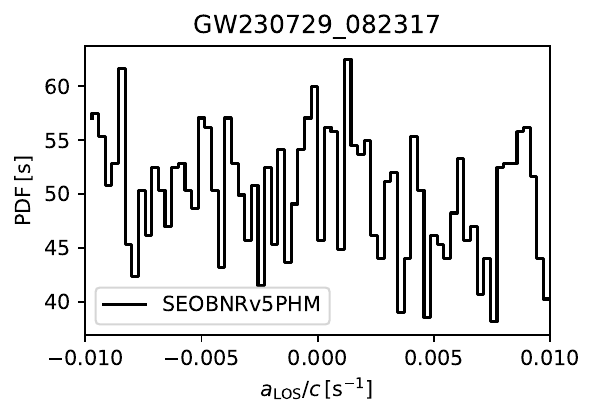}
    \caption{Inferred line-of-sight acceleration $a/c$ of GW230729\_082317 using the parameterized SEOBNRv5PHM waveform model, with a uniform prior over $[-0.01, 0.01]\,\mathrm{s}^{-1}$, where $c$ denotes the speed of light.}
    \label{fig:acc}
\end{figure}

\begin{figure}[htbp]
    \centering
    \includegraphics[width=0.5\linewidth]{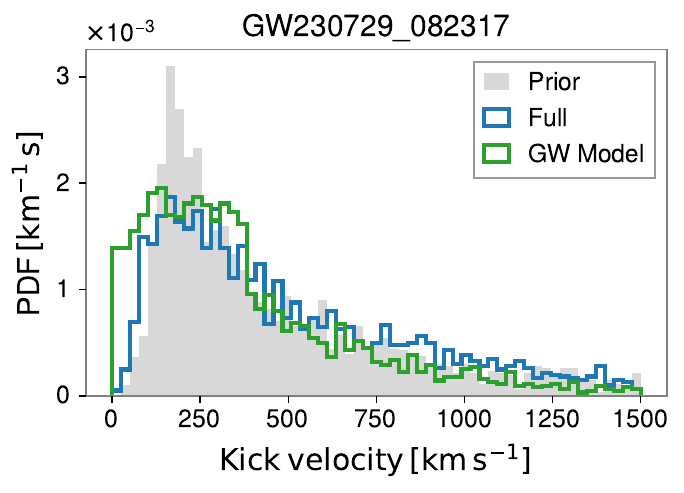}
    \caption{Distribution of recoil velocities for the GW230729\_082317 remnant. The gray shaded histograms represent the kick velocity derived from the assumed priors on the binary source parameters \citep{2026arXiv260404546I}. The dark blue and green curves correspond to results obtained using the gwModel and NRSur7dq4Remnant methods, respectively.}
    \label{fig:kick-1}
\end{figure}

Dynamical interactions are expected to leave characteristic imprints such as orbital eccentricity and line-of-sight acceleration. The detection of non-zero eccentricity or acceleration in GW230729\_082317 would therefore strongly support a hierarchical triple merger origin. We performed dedicated searches for both signatures. The eccentricity analysis employed the non-precessing, higher-order mode waveform model SEOBNRv5EHM \citep{2025PhRvD.112d4038G}. For the acceleration search, the total GW was parameterized as $\Psi(f)=\Psi_0(f)+\Delta \Psi(f)$, where $\Psi_0(f)$ is the phase of an isolated binary and the phase correction $\Delta \Psi(f)$ induced by line-of-sight acceleration follows the prescription of  \citet{2023ApJ...954..105V} up to $3.5$ PN order. As shown in Figures~\ref{fig:ecc} and \ref{fig:acc}, neither signature was significantly detected. Our analysis of GW230729\_082317 finds no significant evidence for line-of-sight acceleration, in agreement with previous analyses \citep{2026arXiv260625304R}. Nevertheless, the absence of measurable eccentricity or acceleration does not rule out a dynamical origin, as many dynamically assembled binaries merger without leaving observable imprints in the frequency band of ground detectors. In a scenario where three black holes undergo hierarchical mergers, the recoil velocity of the merger remnant must be relatively low. Otherwise, it would escape the dense host environment or disrupt the triple system in which it formed. The inferred recoil velocity, presented in Figure~\ref{fig:kick-1}, is obtained using the gwModel \citep{2026PhRvD.113j4017I} and NRSur7dq4Remnant \citep{2026arXiv260404546I} methods. For globular clusters with masses between $10^5~M_\odot$ and $10^6~M_\odot$, the escape speed ranges from $10~{\rm km/s} \le v_{\rm esc} \le 60~{\rm km/s}$ \citep{2002ApJ...568L..23G}. In an AGN disk, the typical orbital velocity is $20,000~{\rm km/s}$, far exceeding the inferred recoil velocities\citep{2019PhRvL.123r1101Y}. If the GW230729\_082317 merger took place in an AGN disk, its remnant is expected to remain bound to the environment and can undergo a subsequent merger. Conversely, had it occurred in a globular cluster, the remnant might have been ejected due to the lower escape speed.


\subsection{Joint Bayesian Analysis}

To statistically assess the association between the two events, we performed a joint parameter estimation (JPE) analysis alongside separate Bayesian analyses for each event. The JPE incorporates the parameter correlations expected in a hierarchical triple merger scenario \citep{2024ApJ...965...80G}. The Bayesian evidences were evaluated independently for the hierarchical triple merger hypothesis and the null (no-association) hypothesis. The expression is as follows:
\begin{eqnarray}
    p(d_1, d_2|\mathcal{H}_H) = \int p(d_1|\boldsymbol{\theta_1}, \boldsymbol{\theta_{\rm ext}}) p(d_2|\boldsymbol{\theta_2}, \boldsymbol{\theta_{\rm ext}}, m_f, a_f) p(\boldsymbol{\theta_1}, \boldsymbol{\theta_2}, \boldsymbol{\theta_{\rm ext}}) \mbox{d}\boldsymbol{\theta_1}\mbox{d}\boldsymbol{\theta_2}\mbox{d}\boldsymbol{\theta_{\rm ext}} \mbox{, }
    \label{eq:evi1}
\end{eqnarray}
\begin{eqnarray}
    p(d_1, d_2|\mathcal{H}_N) &=& \int p(d_1|\boldsymbol{\theta_1}) p(d_2|\boldsymbol{\theta_2}) p(\boldsymbol{\theta_1}) p(\boldsymbol{\theta_2}) \mbox{d}\boldsymbol{\theta_1}\mbox{d}\boldsymbol{\theta_2} \\ \nonumber
    &=& \int p(d_1|\boldsymbol{\theta_1}) p(\boldsymbol{\theta_1}) \mbox{d}\boldsymbol{\theta_1} \times \int p(d_2|\boldsymbol{\theta_2}) p(\boldsymbol{\theta_2}) \mbox{d}\boldsymbol{\theta_2} \mbox{,}
    \label{eq:evi2}
\end{eqnarray}
where $m_f$ and $a_f$ denote mass and spin of the remnant from the first GW event. The Bayes factor between the hierarchical triple merger hypothesis $\mathcal{H}_H$ and the null hypothesis $\mathcal{H}_N$ is
\begin{eqnarray}
    \mathcal{O}_{\mathcal{H}_N}^{\mathcal{H}_H} = \frac{p(d_1, d_2|\mathcal{H}_H)}{p(d_1, d_2|\mathcal{H}_N)} \mbox{.}
    \label{eq:jpe}
\end{eqnarray}

\begin{figure}[htbp]
    \centering
    \includegraphics[width=0.99\linewidth]{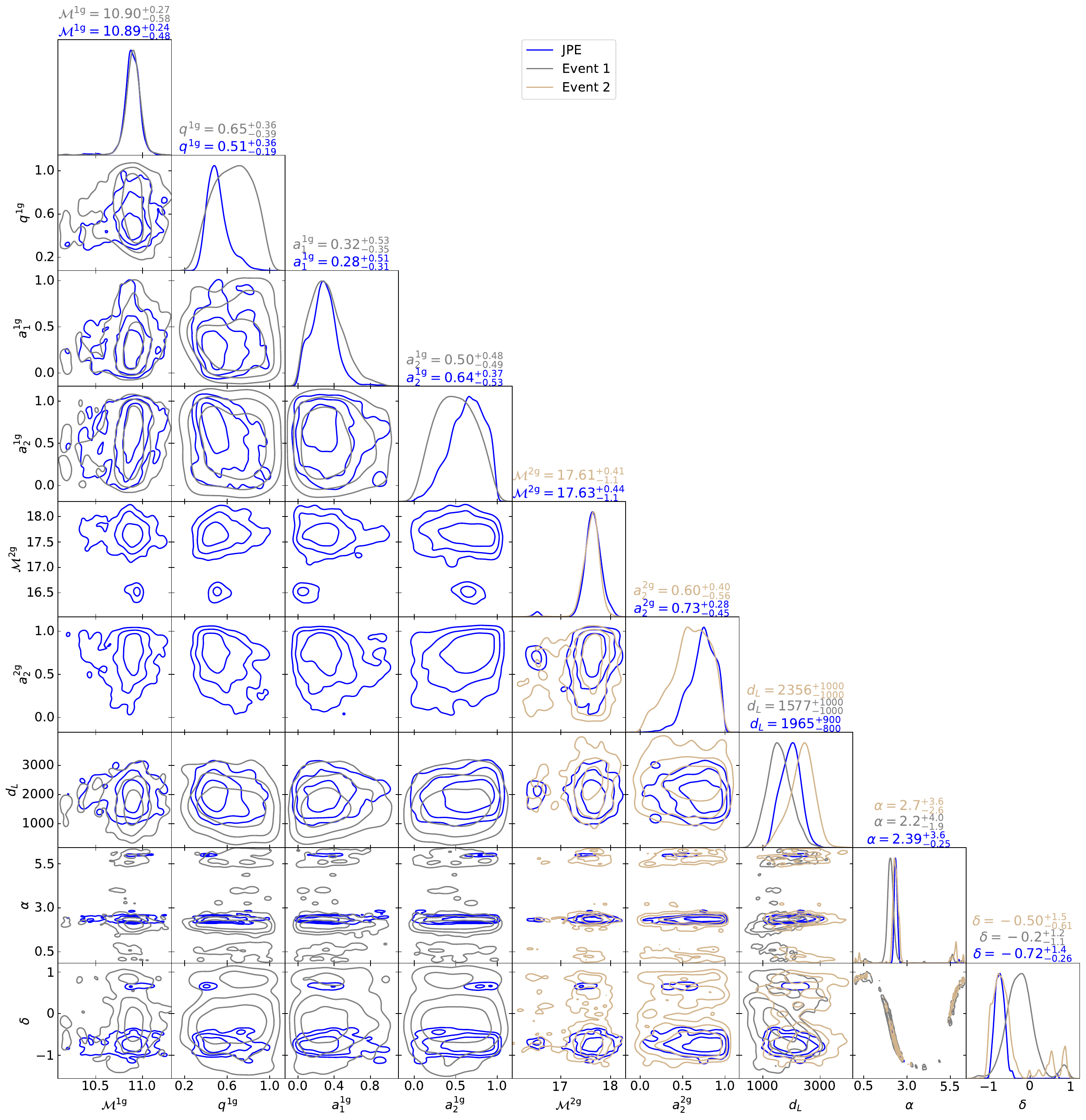}
    \caption{Posterior distributions of the key parameters obtained from the JPE analysis. The figure includes both intrinsic parameters, such as the chirp masses $\mathcal{M}$, mass ratios $q$ and spins, and extrinsic parameters, including the luminosity distance $d_{\rm L}$ and the sky location, characterized by the right ascension $\alpha$ and declination $\delta$. The superscripts ``1g" and ``2g" indicate the first and second mergers, respectively.}
    \label{fig:jpe}
\end{figure}

The results of the joint and individual Bayesian analyses, based on the SEOBNRv5PHM waveform model \citep{2023PhRvD.108l4037R}, are summarized in Figure~\ref{fig:jpe}. Compared to the standalone analyses, the JPE approach noticeably tightens the constraints on the intrinsic parameters, including masses and spins. Meanwhile, the posteriors of the extrinsic parameters, including sky location and luminosity distance, are in good agreement between the joint and separate analyses. These findings are consistent with a hierarchical triple merger interpretation. According to Equation~(\ref{eq:jpe}), the Bayes factor in favor of the hierarchical triple merger hypothesis over the null hypothesis is $110$. While this value alone strongly favors the association scenario, the hypothesis is no longer supported once the prior knowledge of the event rate is incorporated. For example, if hierarchical triple merger constitute only $0.1\%$ of all detected binary black hole mergers \citep{2023MNRAS.523.4113T}, the prior odds in favor of $\mathcal{H}_H$ are $0.001$. The resulting posterior odds ratio is then about $0.1$, indicating that the null hypothesis is favored once the rarity of such systems is accounted for.

The data for GW230729\_082317 and GW231118\_005626 suggest a correlation between these events, as quantified by the Bayes factor. However, given the intrinsic rarity of hierarchical triple merger, no definitive statistical conclusion can be drawn. As discussed in Section~\ref{sec:form}, certain dynamical processes can naturally account for such a pair of mergers, and therefore the absence of measurable eccentricity and line-of-sight acceleration does not undermine the hierarchical triple merger hypothesis.

\section{Discussion on Formation Pathways}\label{sec:form}

Hierarchical triple merger can originate through either primordial triple evolution or dynamical interactions. Given the high multiplicity fraction of massive stars, one possible interpretation is that GW230729\_082317 and GW231118\_005626 originated from a primordial hierarchical triple formed during stellar evolution. Alternatively, hierarchical triple systems may be assembled dynamically through repeated few-body interactions in environments such as globular clusters, nuclear star clusters and active galactic nucleus (AGN) disks. In the following, we discuss these possible formation channels. We first introduce the theoretical framework governing the secular evolution of hierarchical triple systems, and then briefly review the simulation of scattering dynamics including the effects of GW emission.

A hierarchical triple system consists of an inner binary orbited by a distant tertiary companion, with the inner and outer orbits well separated so that the system remains dynamically stable over long timescales. Because the tertiary is relatively distant, studies of hierarchical triple typically focus on the evolution of the inner binary under the tertiary perturbation, while neglecting the evolution of the outer orbit. The degree of hierarchy determines whether a double-orbit average or only an inner-orbit average is appropriate. The equations of motion governing hierarchical triple systems are given by prior work \citep{2018ApJ...863...68L,2019ApJ...881...41L}.
When the mutual inclination between the inner and outer orbit planes lies in the range $39^\circ-141^\circ$, the system undergoes LK oscillations, characterized by periodic variations in eccentricity and inclination. In the quadrupole approximation, inclination near $90^\circ$ can drive the eccentricity toward unity. As the octupole parameter $\epsilon_{\rm oct} \equiv a_1e_2(m_1-m_2)/(a_2(1-e_2^2)(m_1+m_2))$ increases, the octupole order term becomes dynamically important \citep{2013MNRAS.431.2155N}. In this regime, extreme eccentricities can be attained even from modest inclinations, alongside a range of qualitatively new dynamical phenomena \citep{2015MNRAS.447..747L}. The stability condition of hierarchical triple is \citet{2001MNRAS.321..398M}
\begin{eqnarray}
    \frac{a_{\rm out}}{a} > 2.8\left( 1 + \frac{m_3}{m_{12}} \right)^{2/5} \frac{\left( 1 + e_{\rm out} \right)^{2/5}}{\left( 1 - e_{\rm out} \right)^{6/5}} \left( 1 - \frac{0.3I_0}{180^\circ} \right) \mbox{, }
    \label{eq:stab}
\end{eqnarray}
here $I_0$ is the mutual inclination angle between the inner and outer orbit planes. The stability of hierarchical triple system increases as the outer orbit widens and the inner orbit tightens. Consequently, if GW230729\_082317 and GW231118\_005626 indeed originated from a primordial hierarchical triple, the system must have maintained long-term dynamical stability. 

Analytical theories are applicable only when the tertiary can be treated as a weak perturbation. For more tightly bound configurations, few-body numerical simulations are required to study the three-body dynamics. Numerous studies have employed few-body simulations within the framework of Newtonian mechanics\citep{1983ApJ...268..319H,2004MNRAS.352....1F,2012A&A...537A.128R,2018MNRAS.480L..58A,2023IAUS..362..404T}. To capture general relativistic effects, the widely adopted approach is to include post-Newtonian (PN) corrections to the Newtonian force in the equations of motion. This is particularly important for scattering simulations, where PN  terms enable the study of GW capture and the assembly of inspiraling bound multiples. Our simulations are performed with the SpaceHub code, which integrates PN-corrected accelerations and is well suited for both hierarchical triple evolution and few-body scattering \citep{SpaceHub2021}. The PN approximation to the gravitational force follows the standard prescription given as \citet{1985AIHPA..43..107D,1989racm.book.....S}
\begin{eqnarray}
    F_{\rm GR} = c^{-2}F_{\rm 1PN} + c^{-4}F_{\rm 2PN} + c^{-5}F_{\rm 2.5PN} \mbox{,}
\end{eqnarray}
where the $c^{-2}F_{\rm 1PN}$ terms accounts for periastron precession, the $c^{-4}F_{\rm 2PN}$ term provides higher order corrections, and the $c^{-5}F_{\rm 2.5PN}$ term corresponds to the gravitational radiation reaction. The latter induces energy and angular momentum losses through GW emission. For an isolated binary dominated by GW emission, the secular evolution of the orbital elements follows the well known Peters formulas \citep{1963PhRv..131..435P}, which are given by
\begin{eqnarray}
    \left. \frac{da}{dt} \right|_{\rm GW} &=& -\frac{64}{5}\frac{G^3}{c^5}\frac{m_1m_2(m_1+m_2)}{a^3(1-e^2)^{7/2}} \left( 1 + \frac{73}{24}e^2 + \frac{37}{96}e^4 \right) \mbox{,} \label{eq:peter-1}\\
    \left. \frac{de}{dt} \right|_{\rm GW} &=& -\frac{304}{15} \frac{G^3}{c^5} \frac{m_1m_2(m_1+m_2)}{a^4(1-e^2)^{5/2}} e\left( 1 + \frac{121}{304}e^2 \right) \mbox{.} 
    \label{eq:peter-2}
\end{eqnarray}
Under the interpretation that the remnant of GW230729\_082317 constitutes one component of the GW231118\_005626 binary, applying Equation~(\ref{eq:peter-1}) and (\ref{eq:peter-2}) yields an estimated initial semi-major axis of approximately $3\times 10^{-4}$AU for GW231118\_005626. At this separation, which exceeds $10^3$ Schwarzschild radii, the tertiary induces a line-of-sight acceleration of $a/c \approx 0.002 \,{\rm s^{-1}}$, a value that is not ruled out by the posterior distribution shown in Figure~\ref{fig:acc}.

\subsection{Primordial Triple}

To evaluate the plausibility of a primordial triple interpretation, we construct a hierarchical triple system adopting the mode values of the posterior mass distributions for the two GW events: $m_1=13.3~M_\odot$, $m_2=7.5~M_\odot$ and $m_3=11~M_\odot$. We then use the SpaceHub code to test whether this hierarchical triple is consistent with the GW data.

\begin{figure}[htbp]
    \centering
    \includegraphics[width=0.99\linewidth]{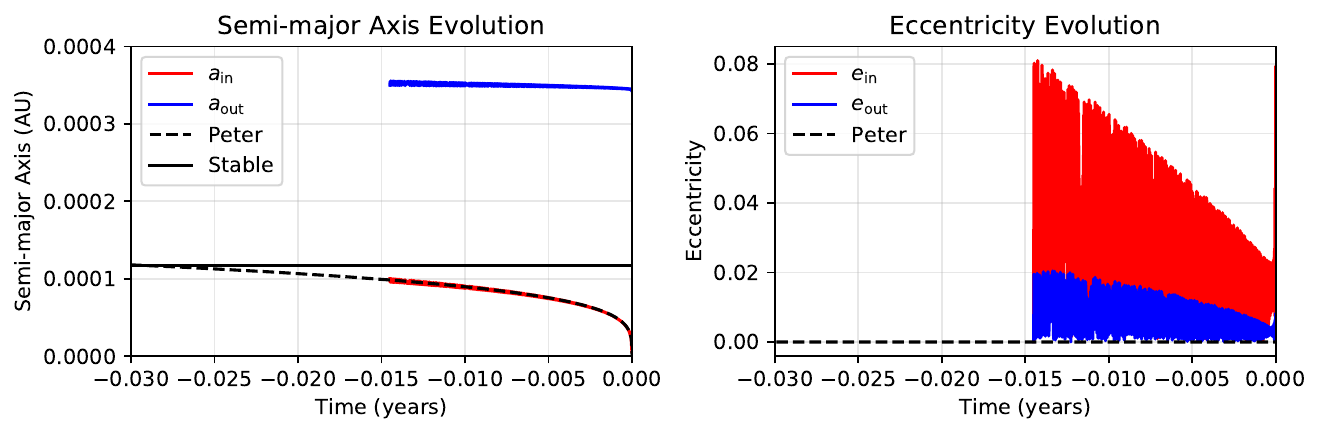}
    \caption{The figure shows the temporal evolution of the semi-major axes in the triple system. The inner binary consists of two bodies with masses of $13.3 M_\odot$ and $7.5 M_\odot$, and its semi-major axis is $10^{-4}$ AU. The tertiary body has a mass of $11 M_\odot$, and the outer semi-major axis is $3.5 \times 10^{-4}$ AU. These mass and outer semi-major axis parameters are inferred from GW data. The oscillations visible in the figure arise from period variations and numerical errors under these conditions.}
    \label{fig:hier_orb}
\end{figure}

Simulations incorporating stability criteria demonstrate that a primordial triple origin is not viable. If the system formed from a stable primordial triple, it would be expected to remain dynamically stable up to point merger. However, both the numerical simulations and the Peters relation displayed in Figure~\ref{fig:hier_orb} indicate that the triple can only maintain stability for approximately $0.03$ years in the vicinity of the merger. The dashed line in Figure~\ref{fig:hier_orb} shows the semi-major axis decay predicted by the Peters formulas, which agrees well with the numerical results. As the ratio of the outer to inner semi-major axis passes through the unstable region, the eccentricity and the inner semi-major axis undergo significant oscillations during their decay. It is therefore unlikely that GW230729\_082317 and GW231118\_005626 arose from sequential mergers in a long-lived, stable hierarchical triple system.

When the outer orbit has a large eccentricity, the assumptions underlying both perturbation theory and orbital averaging break down, even in the presence of a substantial hierarchy between the inner and outer semi-major axes \citep{2026MNRAS.549ag944G}. This regime lies beyond the scope of the present discussion. In this work, we consider a coplanar configuration. The advantage of this coplanar setup is that the angular momentum brought in by the incoming single star can substantially reduce the orbital angular momentum of the initial target binary, thereby yielding a shorter GW merger timescale. These coplanar systems predominantly contribute to prompt second generation mergers \citep{2019MNRAS.482...30S}.

Dynamical hardening of a primordial triple may produce the observed short time interval double merger. This channel, which belongs to the category of higher order scattering, is discussed next.

\subsection{Binary-Single Interaction}

The rich variety of scattering processes can give rise to unusual merger configurations not typically expected from isolated binary evolution.
Hierarchical triples can be assembled through several dynamical scattering channels, including binary-single scattering (only intermediate triples states), binary-binary scattering, triple scattering, and higher order multiple interactions. In the following, we discuss each channel briefly.

In purely Newtonian mechanics, the formation of a stable triple through a single scattering event is extremely difficult. Resonant binary-single encounters may temporarily assemble unstable triple that eventually decay into an eccentric binary. Compared to the full initial hyperparameter space, the probability of forming a permanently stable triple system through binary-single scattering is so small that it is considered measure zero in Newtonian theory \citep{JeanChazy1929,1983ApJ...268..319H}. Despite this, binary-single scattering remains a more efficient channel for producing eccentric GW sources than single-single scattering \citep{2014ApJ...784...71S}.

Nevertheless, stable triple systems can form through either dissipative processes or via repeated dynamical interactions in dense stellar environment. Tidal dissipation efficiently removes energy and facilitates triple formation in stellar systems. However, this mechanism does not operate for black holes \citep{1989ApJ...341..175B}. In binary-single encounters, bodies with internal structure, such as stars with extended gaseous envelopes or white dwarfs, can alter the chaotic resonant evolution of the transient triple, thereby enhancing the probability of capture over exchange \citep{2017ApJ...846...36S,2018MNRAS.481.5436S}. For black holes, GW emission provides the only intrinsic dissipative mechanism during close encounters. However, GW capture requires nearly horizon-scale pericenter passages and low encounter velocities, and is most effective for near-equal mass ratio ($m_2 \approx m_3$), whereas mass asymmetry preferentially leads to non-capture scattering or exchanges \citep{2025A&A...697A.118R}. Consequently, the direct formation of a stable hierarchical triple through few body scattering is expected to be highly inefficient, with exchanges, ionization, and prompt mergers constituting the dominant outcomes, as demonstrated in the no GR case \citep{2016MNRAS.456.4219A}. Previous work \citep{2019MNRAS.482...30S} provides a detailed account of triple formation via equal-mass binary-single scattering. We briefly address the unequal-mass regime below.

\begin{figure}[htbp]
    \centering
    \includegraphics[width=0.49\linewidth]{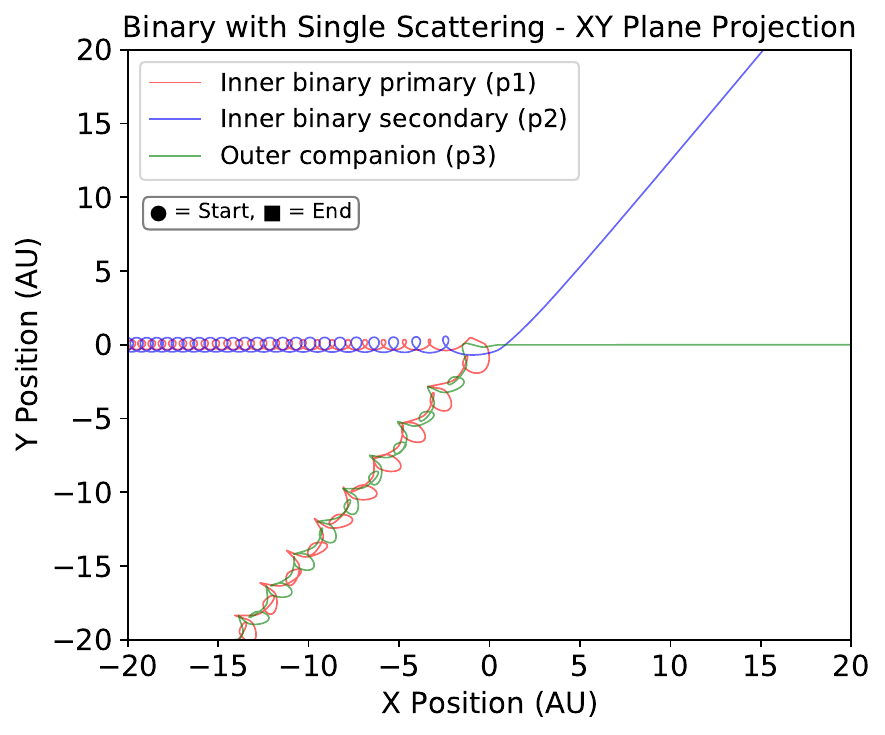}
    \includegraphics[width=0.49\linewidth]{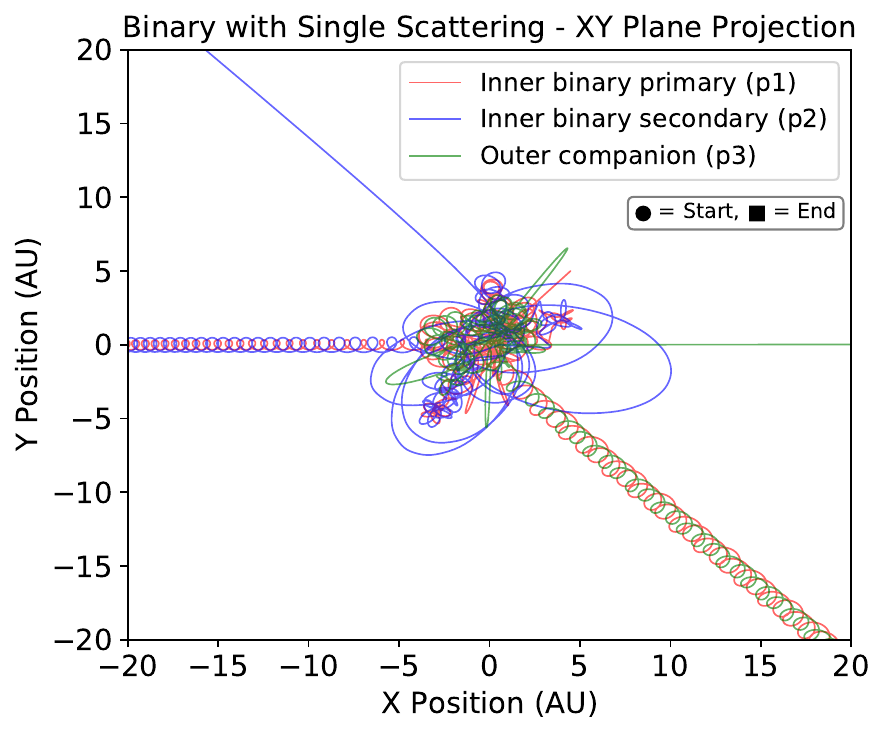}
    \caption{The figure illustrates a binary-single scattering process. The initial binary consists of an $11 M_\odot$ primary and a $7.5 M_\odot$ secondary and is perturbed by a $13.3 M_\odot$ tertiary intruder. As a result of the interaction, object $1$ is ejected, while the remaining two bodies form a bound binary. Compared with the case shown in the left panel, the case shown in the right panel has a lower initial velocity, a larger impact parameter, and a different angle of incidence.}
    \label{fig:b2s}
\end{figure}

We proceed by characterizing the dynamical conditions for strong scattering in single-binary configurations. The strength of the gravitational interaction is governed by the impact parameter $b$ and the velocity $v_\infty$ of the third body with respect to the target binary. In the two-body scattering model, the impact parameter is directly related to the distance of closest approach $r_{\rm min}$ by
\begin{eqnarray}
    b = r_{\rm min} \sqrt{1 + \frac{2Gm_{\rm tot}}{r_{\rm min}v_{\infty}^2}} \mbox{.}
    \label{eq:rmin}
\end{eqnarray}
Its ratio to the semi-major axis of the target binary serves as a key parameter for quantifying the strength of the interaction. Whether the geometric cross-section $\pi r_{\rm IC}^2$ or the gravitational focusing term $\pi r_{\rm IC}\frac{2Gm_{\rm tot}}{v_\infty^2}$ dominates depends on the ratio of the binding energy of the target binary to the kinetic energy of the incoming object. Here, $r_{\rm IC}$ is defined as $r_{\rm IC} = \frac{m_1a_0}{m_1+m_2}$, with $m_1$ and $m_2$ denoting the component masses of the binary (assuming $m_1 > m_2$) \citep{2014ApJ...784...71S}. Here, we simulate scattering in the gravitational focusing regime. The characteristic velocity $v_c$ corresponds to the threshold at which the total energy of the system equals zero. For binary-single scattering, $v_c$ is \citep{1975MNRAS.173..729H,1983ApJ...268..319H}
\begin{eqnarray}
    v_{\rm c, bin-sin}^2 = \frac{Gm_1m_2(m_1+m_2+m_3)}{a_0m_3(m_1+m_2)} \mbox{.}
\end{eqnarray}
The inclusion of Post-Newtonian force corrections suppressed the formation of transient hierarchical configurations, as the premature of the inner binary truncates the resonant interactions.
The energy radiated in a single close parabolic encounter with isolated-body system is given by \citep{2010PhRvD..82j7501B,2012PhRvD..86d4017D}
\begin{eqnarray}
    \Delta E_{\rm parabolic} = \frac{64\pi}{5} \frac{G^3}{c^5} \frac{m_1^2m_2^2}{r_p^2} \omega_c \frac{1+(73/24)e^2+(37/96)e^4}{(1+e)^{7/2}} \mbox{,}
\end{eqnarray}
where $\omega_c=\sqrt{G(m_1+m_2)/r_p^3}$ is the angular frequency of a circular orbit of radius $r_p$. Under specific conditions, GW dissipation during binary-single scattering can produce a transient bound triple system. As shown in the right panel of the Figure~\ref{fig:b2s}, the three body form a transient triple system for about $20$ years. However, such scenarios are short-lived, typically results in eccentric mergers, and are not relevant to GW230729\_082317 and GW231118\_005626.

In the hard-binary limit, the total energy of the system is negative, and the binary survives for a few to several thousand initial orbital periods. These systems constitute the primary focus of this study \citep{1983ApJ...268..319H,1993ApJ...403..256H}. The simulation shown in Figure~\ref{fig:b2s}, with an initial velocity $v_\infty=0.1~v_c$, represents a typical case of strong dynamical scattering. Because of challenges in accurately resolving few-body interactions numerically, such as insufficient precision in estimating tidal energy losses for each orbit and the treatment of angular momentum dissipation \citep{2018MNRAS.481.5436S}, we restrict our focus to a schematic proof-of-concept. A fully detailed treatment is deferred to a subsequent study. The time interval between the two GW mergers is minimized in binary-single interactions when the total angular momentum of the system approaches zero. A natural astrophysical setting for such interactions is the accretion disk of an AGN, where the first merger tends to occur on an eccentric orbit \citep{2019MNRAS.482...30S}. The initial angular momentum of a binary-single system is closely linked to the final observable signature of a double GW merger. The inclusion of single and binary scattering in AGN disks not only facilitates the formation of stable triple systems, but also enables these gas-assisted triples to subsequently produce binary gravitational-wave mergers \citep{2022Natur.603..237S,2025MNRAS.539.1501R,2025ApJ...983..114W,2025ApJ...993..139W}.

\subsection{Binary-Binary Interaction}

\begin{figure}[htbp]
    \centering
    \includegraphics[width=0.6\linewidth]{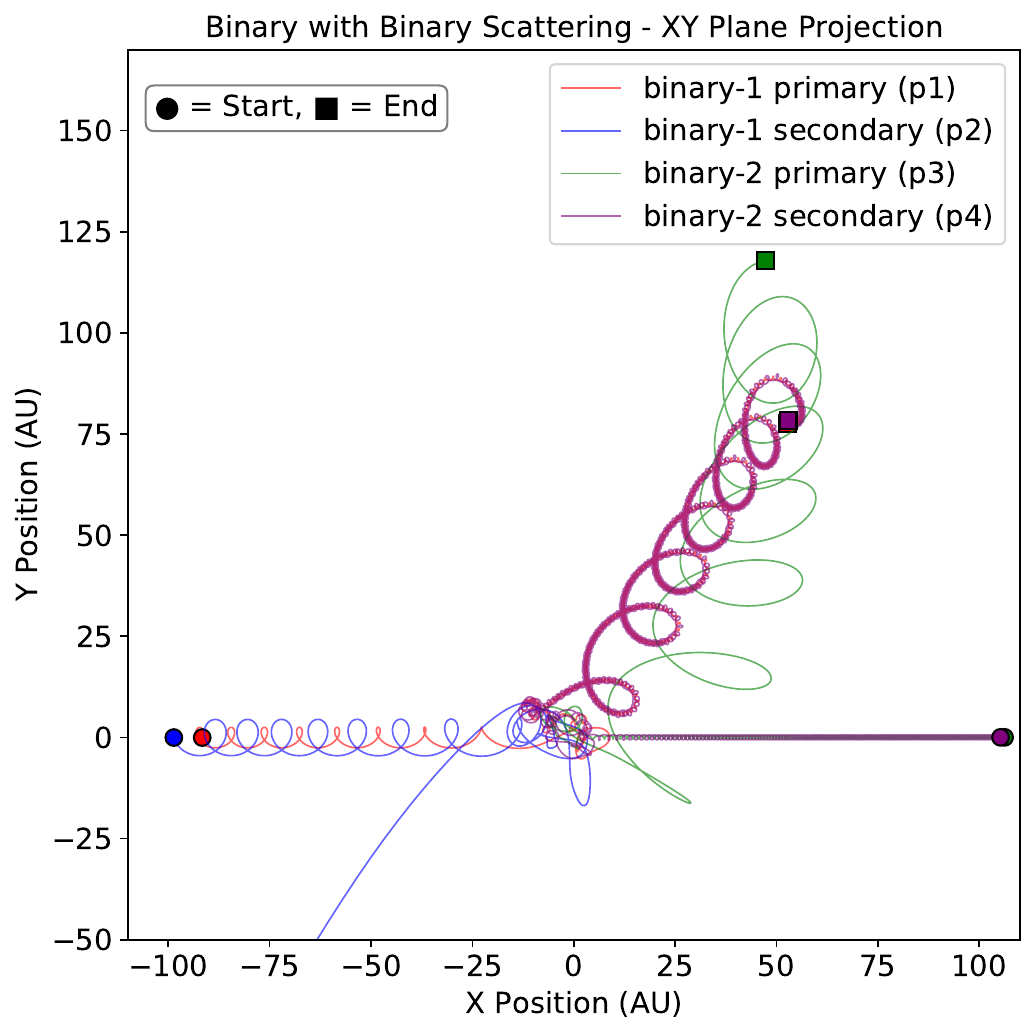}
    \caption{The figure present an example of a scattering process that could give rise to the GW events GW230729\_082317 and GW231118\_005626. The four participating particles have masses (ordered numerically) of $13.3 M_\odot$, $7.5 M_\odot$, $11 M_\odot$ and $7.5 M_\odot$. In this interaction, particle $4$ is ejected, leaving the remaining three bodies to form a bound triple system.}
    \label{fig:b2b}
\end{figure}

Despite being 10-100 times less frequent than binary-single encounters in globular clusters, binary-binary interactions are more likely to form triple systems, owing to their prolonged durations and richer dynamical complexity \citep{2019ApJ...871...91Z}. The probability of a binary-binary encounter leading to a GW capture is 3-5 times higher than that of a binary-single encounter. Binary-binary black hole encounters in cluster cores are therefore crucial to the formation of triples systems and for GW captures events involving three bodies. Reference \citet{2016MNRAS.456.4219A}, which did not include PN effects, demonstrated that binary-binary scattering constitutes an important channel for the formation of triple systems. These dynamically assembled triples are generally compact, with outer to inner semi-major axis ratios in the range $\sim 3-100$, and exhibit an approximately isotropic mutual inclination distribution. Their outer orbital eccentricities are consistent with thermal distribution. Consequently, a substantial fraction of those system undergo efficient LK oscillations, which drive the inner binary to extreme eccentricities and thereby promote rapid mergers via GW emission. The possible outcomes of binary-binary scattering include GW capture, stable triples formation, and the flyby of binary-binary or ionization configurations. Among these, GW capture arises as a direct consequence of the 2.5PN radiation reaction force, whereas stable triple systems emerge from purely Newtonian dynamics. The critical velocity $v_{\rm c}$ for binary-binary scattering, is given by
\begin{eqnarray}
    v_{\rm c, bin-bin}^2 = \left( \frac{m_1m_2}{a_1} + \frac{m_3m_4}{a_2}\right) \frac{G(m_1+m_2+m_3+m_4)}{(m_1+m_2)(m_3+m_4)} \mbox{,}
\end{eqnarray}
where the two interacting binaries consist of masses $m_1,m_2$ with semi-major axis $a_1$, and $m_3,m_4$ with semi-major axis $a_2$, respectively. Consistent with the properties of GW230729\_082317 and GW231118\_005626, the scattering-produced triples are highly compact, typically exhibiting an outer-to-inner semi-major axis ratio of $\sim 10$. As illustrated in the Figure~\ref{fig:b2b}, the scattering between two initially unbound stars triggers a triple interaction, with the mass drawn from their posterior distributions. The eccentricity distribution of dynamically formed triples exhibits two components. One is a high-eccentricity component that approximately follows a thermal distribution; the other is a peak located at the initial eccentricity of the inner binary. For systems with initially circular inner orbits, this peak appears at zero. The formation rate of bound triples via triple-single scattering exceeds that from binary-binary scattering by several orders of magnitude, both in globular clusters and field stars. Given the substantial observed population of triples \citep{2025PASP..137i4201S}, triple scattering emerges as a promising channel for forming compact triple systems. Scattering involving hierarchical triples can similarly produce new compact hierarchical triples, a pathway we do not elaborate further here.

Under favorable conditions, dynamical scattering of binaries can produce hierarchical triples whose successive mergers yield event pairs similar to GW230729\_082317 and GW231118\_005626. The low recoil velocity inferred for GW230729\_082317 suggest that an AGN disk is the most probable environment for this process, as migration traps within such disks provide natural sites for repeated mergers (see also \citet{2026ApJ...999..127L,2026arXiv260701121L}).

\section{Constraints on Events Rate} \label{sec:rate}

The LVK detector network has completed four observing runs since 2015. They are as follows, O1(lasting approximately four months), O2(roughly nine months), O3(about eleven months, conducted in two segments), and O4 (spanning approximately two and a half years and subdivided into O4a, O4b and O4c). An interim run is planned for late 2026, followed by the fifth observing run (O5) \citep{Arnaud:2026hzn}. In this work, we estimated the hierarchical triple merger rate using the observed number of BBH merger events. Assuming that merger events constitute a Poisson point process and are uniformly distributed in comoving volume, the probability of observing $k$ events is
\begin{eqnarray}
    P(k | \lambda) = \frac{\lambda^k e^{-\lambda}}{k!} \mbox{, }
\end{eqnarray}
where $\lambda$ denotes the expected number of detected events. Following the statistical framework developed by \citet{2020MNRAS.498L..46V}, we analyzed the GWTC-4 dataset to constrain the rate of hierarchical triple mergers. Since the complete posterior samples for GWTC-5.0 have not yet been publicly released, a systematic search for hierarchical triple merger candidates in the latest catalog is deferred to future work. The probability of observing a hierarchical triple merger depends not only on the intrinsic merger rate but also on the observing durations $\Delta_{i}$ and intervening temporal gaps between successive runs $\Delta_{i-i+1}$, the searching spacetime volume of each runs $\Delta t_iC_i$, the number of BBH events identified in each observing phase $n_{i}$, and the underlying delay-time distribution of two mergers. A detailed form of probability is provided in the Appendix~\ref{sec:prob}. 

In the following, we describe the evaluation of the sensitive spacetime volume and the expected number of hierarchical triple mergers.
According to standard count event by LVK \citep{2016ApJ...833L...1A}, the expected number of detected BBH mergers is given by
\begin{eqnarray}
    \lambda = T_{\rm obs} \int dzd\boldsymbol{\theta} \frac{dV_c}{dz} \frac{R(z)}{1+z}P(\boldsymbol{\theta})\epsilon(z,\boldsymbol{\theta}) \label{eq:lambda}\mbox{,}
\end{eqnarray}
where $P(\boldsymbol{\theta})$ denotes the intrinsic distribution of BBH source parameters, and $\epsilon(z,\boldsymbol{\theta})$ is the selection function describing the detection efficiency. For the underlying BBH population, we adopt the BROKEN POWER LAW + 2 PEAKS model, which is the fiducial mass model used in \citet{2026ApJ..1005L..51A}. The distribution of primary mass is modeled as
\begin{eqnarray}
    \pi(m_1|\Lambda) &\propto& \left[ \lambda_0p_{\rm BP}(m_1|\alpha_1,\alpha_2,m_{\rm break},m_{\rm 1,low},m_{\rm high})+\lambda_1\mathcal{N}_{lt}(m_1|\mu_1,\sigma_1,{\rm low=m_{1,low}})\right. \nonumber\\
    &+& \left.(1-\lambda_0-\lambda_1)\mathcal{N}_{lt}(m_1|\mu_2,\sigma_2,{\rm low=m_{1,low}}) \right] \times S(m_1|m_{\rm 1,low},\delta_{\rm m,1}) ,
\end{eqnarray}
where $p_{\rm BP}$ denotes the broken power law component, $\mathcal{N}_{lt}$ denotes a left-truncated Gaussian distribution, and $S$ is the mass smoothing function. The conditional mass ratio distribution is modeled as a power law, 
\begin{equation}
    p_{\rm PL}(q|m_1,\beta_q,m_{\rm 2,low},\delta_{\rm m,2}) \propto q^{\beta_q}S(m_1q|m_{\rm 2,low},\delta_{m,2}) .
\end{equation}
For the spins distribution, we adopt the GAUSSIAN COMPONENT SPINS model. The spin magnitudes and cosine tilt angle are described by
\begin{eqnarray}
    \pi(\chi_i|\mu_\chi,\sigma_\chi) &=& \mathcal{N}_{[0,1]}(\chi_1|\mu_\chi,\sigma_\chi)\mathcal{N}_{[0,1]}(\chi_2|\mu_\chi,\sigma_\chi), \\
    \pi(\cos{\theta_i}|\mu_t,\sigma_t,\zeta) &=& \zeta\mathcal{N}_{[-1,1]}(\cos\theta_1|\mu_t,\sigma_t) \times \mathcal{N}_{[-1,1]}(\cos\theta_2|\mu_t,\sigma_t) + \frac{1-\zeta}{4} .
\end{eqnarray}
Here, $\mathcal{N}_{[a,b]}$ denotes a Gaussian distribution truncated to the interval $[a,b]$. For the redshift evolution, we assume that the merger rate density follows $R(z)=R_0(1+z)^\kappa$. Further details of the population model and the corresponding posterior are provided in \citet{2026ApJ..1005L..51A} \footnote{The population parameters are set to the median values of the posterior distributions provided in \\\texttt{BBHMassSpinRedshift\_BrokenPowerLawTwoPeaks\_GaussianComponentSpins\_PowerLawRedshift.h5}}. Using the injection campaign released for the O1-O4a observing runs \footnote{\url{https://zenodo.org/records/19500064}\\ \url{https://zenodo.org/records/5546676}} \citep{2018CQGra..35n5009T,2021RNAAS...5..220E,2023PhRvD.108d3011E,2025PhRvD.112j2001E}, we estimate the sensitive spacetime volume and thereby evaluate Eq.~(\ref{eq:lambda}). Because current template-based GW searches aim for circular binaries are less sensitive to eccentric systems \citep{2019ApJ...883..149A}, we conservatively account for this reduction in search sensitivity by applying a $50\%$ efficiency loss, following \citet{2020MNRAS.498L..46V}. To calculate the probability of observing hierarchical triple mergers, we model the time delay between the first and second mergers with a delay time distribution, which determines the conditional probability of observing the second merger. We adopt the delay time distribution $P(\Delta<T) \propto (T/t_{\rm max})^\alpha$ proposed by \citet{2019MNRAS.482...30S}.

\begin{figure}[htbp]
    \gridline{
    \fig{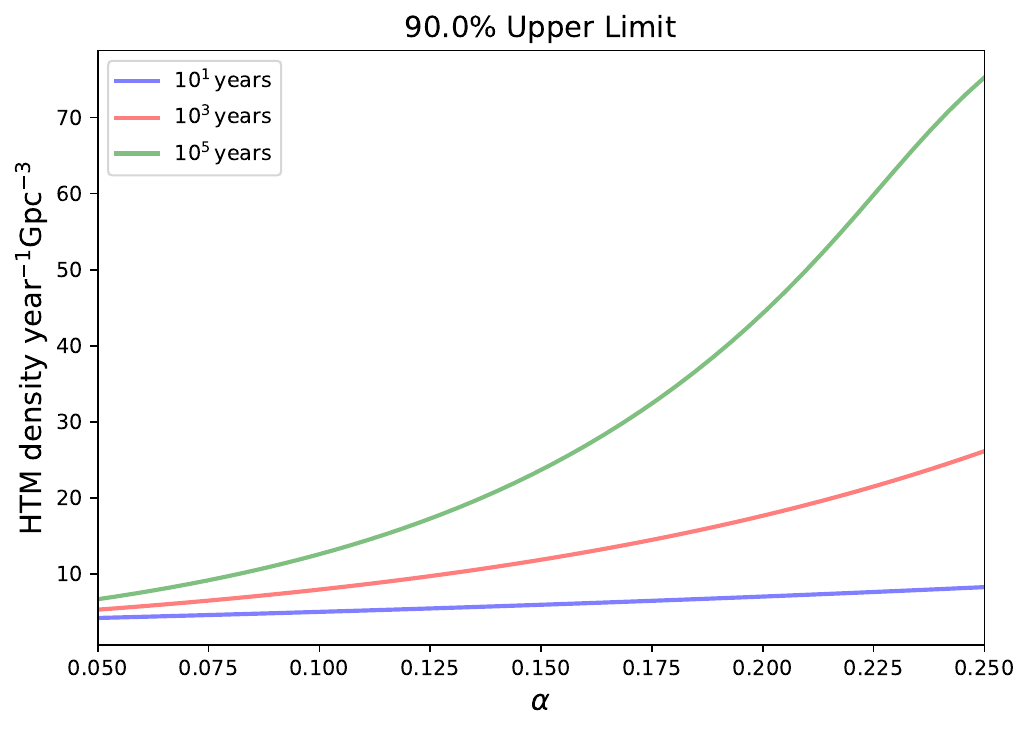}{0.49\textwidth}{(a): The case with no identified hierarchical triple mergers.}
    \fig{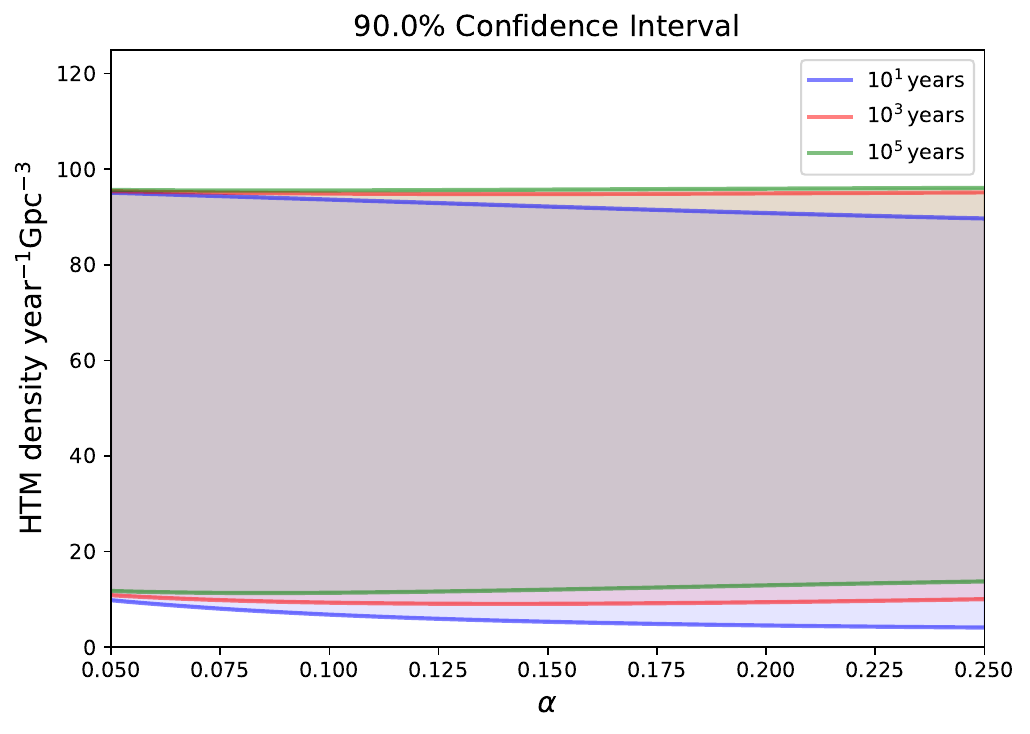}{0.49\textwidth}{(b): The case with one identified hierarchical triple mergers.}
    }
    \caption{Constraints on the local hierarchical triple merger density by GWTC-4 catalog. The constraints are functions of both $t_{\rm max}$ and $\alpha$. We illustrate this parameter dependence by presenting results for three cases, namely $t_{\rm max}=10, 10^3$, and $10^5$ years.}
    \label{fig:rate}
\end{figure}

Then, we can obtained the probability of observing zero or one hierarchical triple merger during the observing period.
The constraints on the local hierarchical triple merger rate density can be derived under specified confidence level. The resulting constraints are shown in Figure~\ref{fig:rate}. The results shown in Figure~\ref{fig:rate}(a) and (b) are obtained by requiring $\mathcal{P}_{\rm null}>0.1$ (Equation~\ref{eq:pn}) and a profile likelihood ratio $-2\ln{(\mathcal{P}_{\rm one}(R_0)/\mathcal{P}_{\rm one}(\hat{R}_0))}<2.706$ (Equation~\ref{eq:po}), respectively, where $\hat{R}_0$ is the value that maximizes $\mathcal{P}_{\rm one}$. In the absence of any observed hierarchical triple merger, the $90\%$ confidence upper limit varies from $\sim 4.2$ to $\sim 75.3$ ${\rm Gpc^{-3}yr^{-1}}$, depending on the choices of $t_{\rm max}$ and $\alpha$. Alternatively, assuming the candidate event pair constitutes a genuine hierarchical triple merger, we obtain a $90\%$ confidence interval for the rate density of $[11.3 \,{\rm Gpc^{-3}yr^{-1}}, 96.1 \,{\rm Gpc^{-3}yr^{-1}}]$ for $t_{\rm max}=10^5$ years. For $t_{\rm max}=10$ years, the interval depends on $\alpha$ and falls within $[4.1\,{\rm Gpc^{-3}yr^{-1}}, 95.1\,{\rm Gpc^{-3}yr^{-1}}]$. For comparison, the GWTC-5.0 catalog yields a local hierarchical merger rate of $0.2-3.11 \, {\rm Gpc^{-3}yr^{-1}}$ at $z=0.2$ \citep{2026arXiv260527226T}. Since hierarchical triple mergers constitute only a subset of the overall hierarchical mergers population, their intrinsic rate is expected to be lower than the total hierarchical merger rate. Therefore, in the scenario where a single hierarchical triple merger is identified, power-law delay time models with large $t_{\rm max}$ and small $\alpha$ become disfavored, and $t_{\rm max}=10^5$ years is excluded under the adopted model. \citet{2020MNRAS.498L..46V} predicted an upper limit of approximately $\sim 11-110 \, {\rm Gpc^{-3} yr^{-1}}$ for the local rate of hierarchical triple mergers, based on the assumption of non-detections throughout the O4 observing run. Our inferred upper limit lies below this predicted range, primarily because a significantly larger number of BBH events have been observed compared to the number anticipated on their study.

\section{Conclusion} \label{sec:concl}

We report the identification of a candidate hierarchical triple merger pair, GW230729\_082317 and GW231118\_005626, from a systematic search of the GWTC-4 catalog. This pair of events shows excellent consistency in both sky location and luminosity distance, while their mass and spin distributions follow the statistical distribution of first and second generation binary black holes. Under the hierarchical triple mergers hypothesis, a joint Bayesian analysis yields a Bayes factor of $110$. However, after accounting for the pow prior probability of such systems, the corresponding posterior odds ratio decreases to $0.1$, indicating that the current data do not provide sufficient evidence to claim the discovery of a hierarchical triple merger event pair. Furthermore, the first merger exhibits no significant signatures of orbital eccentricity or line-of-sight acceleration. Combined with the inferred recoil velocity and the short time interval between the two mergers, these results disfavor a primordial triple origin. Instead, dynamical interactions, particularly binary-binary scattering within the AGN disk, provide a more formation scenario. Based on the identified candidates event pair, we further derive constraints on the local rate density of hierarchical triple mergers. Assuming that no hierarchical triple mergers are observed, the $90\%$ confidence upper limit lies in the range $[4.2, 75.3]\,{\rm Gpc^{-3}yr^{-1}}$, depending on $t_{\rm max}$ and $\alpha$. Alternatively, if the candidate event pair is interpreted as a genuine hierarchical triple merger under a delay model with $t_{\rm max} = 10$ years, the corresponding $90\%$ confidence interval for the rate is $[4.1, 95.1] \, {\rm Gpc^{-3}yr^{-1}}$, with the exact bounds depending on $\alpha$. Based on the currently GW data, there is insufficient evidence to conclude that GW230729\_082317 and GW231118\_005626 constitute a hierarchical triple merger pair. As the analysis of the GWTC-5 catalog progresses and further observing runs expand the binary black hole sample, additional hierarchical triple merger candidates may be identified, enabling more comprehensive tests of the hierarchical triple merger scenario and tighter constraints on its astrophysical rate.

The dynamic simulation presented here neglect black hole spin effects. Given their limited influence on the key observables considered in this work, this simplification does not expected to affect the robustness of our conclusions, although incorporating spin dynamics will provide a more complete physical description and is left for future work \citep{2018ApJ...863....7R}. Although a subpopulation of merged black holes has recently been identified in the  GW data \citep{2026PhRvL.137b1404P}, direct observational evidence for a hierarchical merger chain has not yet been found. The merger signatures of hierarchical triple merger require dedicated follow-up studies. Incorporating physically motivated dynamical model into the rate analysis could further refine the inferred event rate constraints by accounting for the distributions of gas-assisted merger delay times \citep{2025MNRAS.539.1501R,2025ApJ...983..114W,2025ApJ...993..139W} and their dependence on binary properties, including component masses and spins. However, these distributions remain poorly constrained at present, and detailed investigation of their impact is deferred to further work.

\appendix

\section{PROBABILITY OF OBSERVATION}\label{sec:prob}

The probability $\mathcal{P}_{\rm null}$ that no hierarchical triple mergers are identified within the GWTC-4 catalog is
\begin{equation}
\begin{split}
    & \mathcal{P}_{\rm null} = \\
    & \left[ \sum_{i=0}^{n_1} \mathrm{P}(i, \kappa_1 R \Delta t_1 C_1) \frac{1}{\Delta t_1^{i}} \left[ \int_{0}^{\Delta t_1} \left[1 - \left(t_{\rm (1,4a_e)} - t_{\rm (1,4a_b)} + t_{\rm (1,3b_e)} - t_{\rm (1,3b_b)} + t_{\rm (1,3a_e)} - t_{\rm (1,3a_b)} + t_{\rm (1,2_e)} - t_{\rm (1,2_b)} + t_{\rm (1,1_e)}\right)\right]d\tau_1 \right]^{i} \right] \\
    & \times \left[ \sum_{i=0}^{n_2} \mathrm{P}(i, \kappa_2 R \Delta t_2 C_2) \frac{1}{\Delta t_2^{i}} \left[ \int_{0}^{\Delta t_2} \left[1 - \left(t_{\rm (2,4a_e)} - t_{\rm (2,4a_b)} + t_{\rm (2,3b_e)} - t_{\rm (2,3b_b)} + t_{\rm (2,3a_e)} - t_{\rm (2,3a_b)} + t_{\rm (2,2_e)} \right)\right]d\tau_2 \right]^{i} \right] \\
    & \times \left[ \sum_{i=0}^{n_{\rm 3a}} \mathrm{P}(i, \kappa_{\rm 3a} R \Delta t_{\rm 3a} C_{\rm 3a}) \frac{1}{\Delta t_{\rm 3a}^{i}} \left[ \int_{0}^{\Delta t_{\rm 3a}} \left[1 - \left(t_{\rm (3a,4a_e)} - t_{\rm (3a,4a_b)} + t_{\rm (3a,3b_e)} - t_{\rm (3a,3b_b)} + t_{\rm (3a,3a_e)} \right)\right]d\tau_{\rm 3a} \right]^{i} \right] \\
    & \times \left[ \sum_{i=0}^{n_{\rm 3b}} \mathrm{P}(i, \kappa_{\rm 3b} R \Delta t_{\rm 3b} C_{\rm 3b}) \frac{1}{\Delta t_{\rm 3b}^{i}} \left[ \int_{0}^{\Delta t_{\rm 3b}} \left[1 - \left(t_{\rm (3b,4a_e)} - t_{\rm (3b,4a_b)} + t_{\rm (3b,3b_e)} \right)\right]d\tau_{\rm 3b} \right]^{i} \right] \\
    & \times \left[ \sum_{i=0}^{n_{\rm 4a}} \mathrm{P}(i, \kappa_{\rm 4a} R \Delta t_{\rm 4a} C_{\rm 4a}) \frac{1}{\Delta t_{\rm 4a}^{i}} \left[ \int_{0}^{\Delta t_{\rm 4a}} \left[1 - t_{\rm (4a,4a_e)} \right]d\tau_{\rm 4a} \right]^{i} \right]
\mbox{,}\end{split}
\label{eq:pn}
\end{equation}
where $\mathrm{P}(i,k)$ denotes the Poisson probability distribution with mean $k$. The parameters $\kappa, n, R, \Delta t$ and $C$ represent the duty cycle efficiency, the observed number of BBH events, the local hierarchical triple merger rate density, the observing time, and the sensitive search volume, respectively. The remaining parameters are defined as follows:
\begin{eqnarray*}
    t_{\rm (1,1_e)} &=& \kappa_{\rm 1} \left( \frac{\Delta t_1 - \tau}{t_{\rm max}} \right)^\alpha  \\
    t_{\rm (1,2_b)} &=& \kappa_{\rm 2} \left( \frac{\Delta t_1 + \Delta t_{\rm 1-2} - \tau}{t_{\rm max}} \right)^\alpha  \\
    t_{\rm (1,2_e)} &=& \kappa_{\rm 2} \left( \frac{\Delta t_1 + \Delta t_2 + \Delta t_{\rm 1-2} - \tau}{t_{\rm max}} \right)^\alpha  \\
    t_{\rm (1,3a_b)} &=& \kappa_{\rm 3a} \left( \frac{\Delta t_1 + \Delta t_2 + \Delta t_{\rm 1-2} + \Delta t_{\rm 2-3a} - \tau}{t_{\rm max}} \right)^\alpha  \\
    t_{\rm (1,3a_e)} &=& \kappa_{\rm 3a} \left( \frac{\Delta t_1 + \Delta t_2 + \Delta t_{\rm 3a} + \Delta t_{\rm 1-2} + \Delta t_{\rm 2-3a} - \tau}{t_{\rm max}} \right)^\alpha  \\
    t_{\rm (1,3b_b)} &=& \kappa_{\rm 3b} \left( \frac{\Delta t_1 + \Delta t_2 + \Delta t_{\rm 3a} + \Delta t_{\rm 1-2} + \Delta t_{\rm 2-3a} + \Delta t_{\rm 3a-3b} - \tau}{t_{\rm max}} \right)^\alpha  \\
    t_{\rm (1,3b_e)} &=& \kappa_{\rm 3b} \left( \frac{\Delta t_1 + \Delta t_2 + \Delta t_{\rm 3a} + \Delta t_{\rm 3b} + \Delta t_{\rm 1-2} + \Delta t_{\rm 2-3a} + \Delta t_{\rm 3a-3b} - \tau}{t_{\rm max}} \right)^\alpha  \\
    t_{\rm (1,4a_b)} &=& \kappa_{\rm 4a} \left( \frac{\Delta t_1 + \Delta t_2 + \Delta t_{\rm 3a} + \Delta t_{\rm 3b} + \Delta t_{\rm 1-2} + \Delta t_{\rm 2-3a} + \Delta t_{\rm 3a-3b} + \Delta t_{\rm 3b-4a} - \tau}{t_{\rm max}} \right)^\alpha  \\
    t_{\rm (1,4a_e)} &=& \kappa_{\rm 4a} \left( \frac{\Delta t_1 + \Delta t_2 + \Delta t_{\rm 3a} + \Delta t_{\rm 3b} + \Delta t_{\rm 4a} + \Delta t_{\rm 1-2} + \Delta t_{\rm 2-3a} + \Delta t_{\rm 3a-3b} + \Delta t_{\rm 3b-4a} - \tau}{t_{\rm max}} \right)^\alpha \mbox{,}
\end{eqnarray*}
\begin{eqnarray*}
    t_{\rm (2,2_e)} &=& \kappa_{\rm 2} \left( \frac{\Delta t_2 - \tau}{t_{\rm max}} \right)^\alpha  \\
    t_{\rm (2,3a_b)} &=& \kappa_{\rm 3a} \left( \frac{\Delta t_2 + \Delta t_{\rm 2-3a} - \tau}{t_{\rm max}} \right)^\alpha  \\
    t_{\rm (2,3a_e)} &=& \kappa_{\rm 3a} \left( \frac{\Delta t_2 + \Delta t_{\rm 3a} + \Delta t_{\rm 2-3a} - \tau}{t_{\rm max}} \right)^\alpha  \\
    t_{\rm (2,3b_b)} &=& \kappa_{\rm 3b} \left( \frac{\Delta t_2 + \Delta t_{\rm 3a} + \Delta t_{\rm 2-3a} + \Delta t_{\rm 3a-3b} - \tau}{t_{\rm max}} \right)^\alpha  \\
    t_{\rm (2,3b_e)} &=& \kappa_{\rm 3b} \left( \frac{\Delta t_2 + \Delta t_{\rm 3a} + \Delta t_{\rm 3b} + \Delta t_{\rm 2-3a} + \Delta t_{\rm 3a-3b} - \tau}{t_{\rm max}} \right)^\alpha  \\
    t_{\rm (2,4a_b)} &=& \kappa_{\rm 4a} \left( \frac{\Delta t_2 + \Delta t_{\rm 3a} + \Delta t_{\rm 3b} + \Delta t_{\rm 2-3a} + \Delta t_{\rm 3a-3b} + \Delta t_{\rm 3b-4a} - \tau}{t_{\rm max}} \right)^\alpha  \\
    t_{\rm (2,4a_e)} &=& \kappa_{\rm 4a} \left( \frac{\Delta t_2 + \Delta t_{\rm 3a} + \Delta t_{\rm 3b} + \Delta t_{\rm 4a} + \Delta t_{\rm 2-3a} + \Delta t_{\rm 3a-3b} + \Delta t_{\rm 3b-4a} - \tau}{t_{\rm max}} \right)^\alpha \mbox{,}
\end{eqnarray*}
\begin{eqnarray*}
    t_{\rm (3a,3a_e)} &=& \kappa_{\rm 3a} \left( \frac{\Delta t_{\rm 3a} - \tau}{t_{\rm max}} \right)^\alpha  \\
    t_{\rm (3a,3b_b)} &=& \kappa_{\rm 3b} \left( \frac{\Delta t_{\rm 3a} + \Delta t_{\rm 3a-3b} - \tau}{t_{\rm max}} \right)^\alpha  \\
    t_{\rm (3a,3b_e)} &=& \kappa_{\rm 3b} \left( \frac{\Delta t_{\rm 3a} + \Delta t_{\rm 3b} + \Delta t_{\rm 3a-3b} - \tau}{t_{\rm max}} \right)^\alpha  \\
    t_{\rm (3a,4a_b)} &=& \kappa_{\rm 4a} \left( \frac{\Delta t_{\rm 3a} + \Delta t_{\rm 3b} + \Delta t_{\rm 3a-3b} + \Delta t_{\rm 3b-4a} - \tau}{t_{\rm max}} \right)^\alpha  \\
    t_{\rm (3a,4a_e)} &=& \kappa_{\rm 4a} \left( \frac{\Delta t_{\rm 3a} + \Delta t_{\rm 3b} + \Delta t_{\rm 4a} + \Delta t_{\rm 3a-3b} + \Delta t_{\rm 3b-4a} - \tau}{t_{\rm max}} \right)^\alpha \mbox{,}
\end{eqnarray*}
\begin{eqnarray*}
    t_{\rm (3b,3b_e)} &=& \kappa_{\rm 3b} \left( \frac{\Delta t_{\rm 3b} - \tau}{t_{\rm max}} \right)^\alpha  \\
    t_{\rm (3b,4a_b)} &=& \kappa_{\rm 4a} \left( \frac{\Delta t_{\rm 3b} + \Delta t_{\rm 3b-4a} - \tau}{t_{\rm max}} \right)^\alpha  \\
    t_{\rm (3b,4a_e)} &=& \kappa_{\rm 4a} \left( \frac{\Delta t_{\rm 3b} + \Delta t_{\rm 4a} + \Delta t_{\rm 3b-4a} - \tau}{t_{\rm max}} \right)^\alpha \mbox{,}
\end{eqnarray*}
\begin{eqnarray*}
    t_{\rm (4a,4a_e)} &=& \kappa_{\rm 4a} \left( \frac{\Delta t_{\rm 4a} - \tau}{t_{\rm max}} \right)^\alpha \mbox{.}
\end{eqnarray*}
Here, we use the subscripts `b' and `e' to denote the beginning and end of an observing run, respectively. We then use two subscripts to define the parameter $t_{(.,.)}$ associated with the corresponding pairs of epochs. For a first merger occurring at time $\tau$ during O1, $t_{\rm (1, 3ae)}$ denotes the duty cycle weighted cumulative probability that the corresponding second merger occurs before the end of O3a. By the end of O4a, the probability that exactly one complete hierarchical triple merger pair is detected over the full observing period is given by
\begin{equation}
    \mathcal{P}_{\rm one} = P_{\rm 1} + P_{\rm 2} + P_{\rm 3a} + P_{\rm 3b} + P_{\rm 4a} \mbox{.}
    \label{eq:po}
\end{equation}
Here, $P_i=P(E_i)$, where $E_i$ is defined as
\begin{equation}
\begin{split}
E_i = \left\{
\substack{
\text{exactly one complete hierarchical triple merger pair is detected over O1-O4a,} \\
\text{and the first merger of the pair occurs during $O_i$}}
\right\}.
\end{split}
\end{equation}
The second merger of the pair occurs after the first merger and before the end of O4a, either within the same observing run or during a subsequent observing run. Therefore, each term $P_{i}$ represents the probability that exactly one hierarchical triple merger pair is detected over the full O1-O4a observing period, with the first merger occurring during observing run $O_i$. For example, $P_2$ including cases in which the first merger occurs during O2 and the second merger occurs during O2, O3a, O3b or O4a. It is important to note that $P_2$ and $P_{j}$($j\neq2$) are mutually exclusive. The explicit formulation for these terms is provided below:
\begin{equation}
\begin{split}
    & P_{\rm 1} = \\
    & \left[ \sum_{i=1}^{n_1} \mathrm{P}(i, \kappa_1 R \Delta t_1 C_1) \frac{i}{\Delta t_1^{i}} \left[ \int_{0}^{\Delta t_1} \left[1 - \left(t_{\rm (1,4a_e)} - t_{\rm (1,4a_b)} + t_{\rm (1,3b_e)} - t_{\rm (1,3b_b)} + t_{\rm (1,3a_e)} - t_{\rm (1,3a_b)} + t_{\rm (1,2_e)} - t_{\rm (1,2_b)} + t_{\rm (1,1_e)}\right)\right]d\tau_1 \right]^{i-1} \right] \\
    & \times \left[ \sum_{i=0}^{n_2} \mathrm{P}(i, \kappa_2 R \Delta t_2 C_2) \frac{1}{\Delta t_2^{i}} \left[ \int_{0}^{\Delta t_2} \left[1 - \left(t_{\rm (2,4a_e)} - t_{\rm (2,4a_b)} + t_{\rm (2,3b_e)} - t_{\rm (2,3b_b)} + t_{\rm (2,3a_e)} - t_{\rm (2,3a_b)} + t_{\rm (2,2_e)} \right)\right]d\tau_2 \right]^{i} \right] \\
    & \times \left[ \sum_{i=0}^{n_{\rm 3a}} \mathrm{P}(i, \kappa_{\rm 3a} R \Delta t_{\rm 3a} C_{\rm 3a}) \frac{1}{\Delta t_{\rm 3a}^{i}} \left[ \int_{0}^{\Delta t_{\rm 3a}} \left[1 - \left(t_{\rm (3a,4a_e)} - t_{\rm (3a,4a_b)} + t_{\rm (3a,3b_e)} - t_{\rm (3a,3b_b)} + t_{\rm (3a,3a_e)} \right)\right]d\tau_{\rm 3a} \right]^{i} \right] \\
    & \times \left[ \sum_{i=0}^{n_{\rm 3b}} \mathrm{P}(i, \kappa_{\rm 3b} R \Delta t_{\rm 3b} C_{\rm 3b}) \frac{1}{\Delta t_{\rm 3b}^{i}} \left[ \int_{0}^{\Delta t_{\rm 3b}} \left[1 - \left(t_{\rm (3b,4a_e)} - t_{\rm (3b,4a_b)} + t_{\rm (3b,3b_e)} \right)\right]d\tau_{\rm 3b} \right]^{i} \right] \\
    & \times \left[ \sum_{i=0}^{n_{\rm 4a}} \mathrm{P}(i, \kappa_{\rm 4a} R \Delta t_{\rm 4a} C_{\rm 4a}) \frac{1}{\Delta t_{\rm 4a}^{i}} \left[ \int_{0}^{\Delta t_{\rm 4a}} \left[1 - t_{\rm (4a,4a_e)} \right]d\tau_{\rm 4a} \right]^{i} \right] \\
    & \times \left[ \int_{0}^{\Delta t_1} \left(t_{\rm (1,4a_e)} - t_{\rm (1,4a_b)} + t_{\rm (1,3b_e)} - t_{\rm (1,3b_b)} + t_{\rm (1,3a_e)} - t_{\rm (1,3a_b)} + t_{\rm (1,2_e)} - t_{\rm (1,2_b)} + t_{\rm (1,1_e)}\right)d\tau_1 \right]
\mbox{,}\end{split}
\end{equation}
\begin{equation}
\begin{split}
    & P_{\rm 2} = \\
    & \left[ \sum_{i=0}^{n_1} \mathrm{P}(i, \kappa_1 R \Delta t_1 C_1) \frac{1}{\Delta t_1^{i}} \left[ \int_{0}^{\Delta t_1} \left[1 - \left(t_{\rm (1,4a_e)} - t_{\rm (1,4a_b)} + t_{\rm (1,3b_e)} - t_{\rm (1,3b_b)} + t_{\rm (1,3a_e)} - t_{\rm (1,3a_b)} + t_{\rm (1,2_e)} - t_{\rm (1,2_b)} + t_{\rm (1,1_e)}\right)\right]d\tau_1 \right]^{i} \right] \\
    & \times \left[ \sum_{i=1}^{n_2} \mathrm{P}(i, \kappa_2 R \Delta t_2 C_2) \frac{i}{\Delta t_2^{i}} \left[ \int_{0}^{\Delta t_2} \left[1 - \left(t_{\rm (2,4a_e)} - t_{\rm (2,4a_b)} + t_{\rm (2,3b_e)} - t_{\rm (2,3b_b)} + t_{\rm (2,3a_e)} - t_{\rm (2,3a_b)} + t_{\rm (2,2_e)} \right)\right]d\tau_2 \right]^{i-1} \right] \\
    & \times \left[ \sum_{i=0}^{n_{\rm 3a}} \mathrm{P}(i, \kappa_{\rm 3a} R \Delta t_{\rm 3a} C_{\rm 3a}) \frac{1}{\Delta t_{\rm 3a}^{i}} \left[ \int_{0}^{\Delta t_{\rm 3a}} \left[1 - \left(t_{\rm (3a,4a_e)} - t_{\rm (3a,4a_b)} + t_{\rm (3a,3b_e)} - t_{\rm (3a,3b_b)} + t_{\rm (3a,3a_e)} \right)\right]d\tau_{\rm 3a} \right]^{i} \right] \\
    & \times \left[ \sum_{i=0}^{n_{\rm 3b}} \mathrm{P}(i, \kappa_{\rm 3b} R \Delta t_{\rm 3b} C_{\rm 3b}) \frac{1}{\Delta t_{\rm 3b}^{i}} \left[ \int_{0}^{\Delta t_{\rm 3b}} \left[1 - \left(t_{\rm (3b,4a_e)} - t_{\rm (3b,4a_b)} + t_{\rm (3b,3b_e)} \right)\right]d\tau_{\rm 3b} \right]^{i} \right] \\
    & \times \left[ \sum_{i=0}^{n_{\rm 4a}} \mathrm{P}(i, \kappa_{\rm 4a} R \Delta t_{\rm 4a} C_{\rm 4a}) \frac{1}{\Delta t_{\rm 4a}^{i}} \left[ \int_{0}^{\Delta t_{\rm 4a}} \left[1 - t_{\rm (4a,4a_e)} \right]d\tau_{\rm 4a} \right]^{i} \right] \\
    & \times \left[ \int_{0}^{\Delta t_2} \left[t_{\rm (2,4a_e)} - t_{\rm (2,4a_b)} + t_{\rm (2,3b_e)} - t_{\rm (2,3b_b)} + t_{\rm (2,3a_e)} - t_{\rm (2,3a_b)} + t_{\rm (2,2_e)} \right]d\tau_2 \right]
\mbox{,}\end{split}
\end{equation}
\begin{equation}
\begin{split}
    & P_{\rm 3a} = \\
    & \left[ \sum_{i=0}^{n_1} \mathrm{P}(i, \kappa_1 R \Delta t_1 C_1) \frac{1}{\Delta t_1^{i}} \left[ \int_{0}^{\Delta t_1} \left[1 - \left(t_{\rm (1,4a_e)} - t_{\rm (1,4a_b)} + t_{\rm (1,3b_e)} - t_{\rm (1,3b_b)} + t_{\rm (1,3a_e)} - t_{\rm (1,3a_b)} + t_{\rm (1,2_e)} - t_{\rm (1,2_b)} + t_{\rm (1,1_e)}\right)\right]d\tau_1 \right]^{i} \right] \\
    & \times \left[ \sum_{i=0}^{n_2} \mathrm{P}(i, \kappa_2 R \Delta t_2 C_2) \frac{1}{\Delta t_2^{i}} \left[ \int_{0}^{\Delta t_2} \left[1 - \left(t_{\rm (2,4a_e)} - t_{\rm (2,4a_b)} + t_{\rm (2,3b_e)} - t_{\rm (2,3b_b)} + t_{\rm (2,3a_e)} - t_{\rm (2,3a_b)} + t_{\rm (2,2_e)} \right)\right]d\tau_2 \right]^{i} \right] \\
    & \times \left[ \sum_{i=1}^{n_{\rm 3a}} \mathrm{P}(i, \kappa_{\rm 3a} R \Delta t_{\rm 3a} C_{\rm 3a}) \frac{i}{\Delta t_{\rm 3a}^{i}} \left[ \int_{0}^{\Delta t_{\rm 3a}} \left[1 - \left(t_{\rm (3a,4a_e)} - t_{\rm (3a,4a_b)} + t_{\rm (3a,3b_e)} - t_{\rm (3a,3b_b)} + t_{\rm (3a,3a_e)} \right)\right]d\tau_{\rm 3a} \right]^{i-1} \right] \\
    & \times \left[ \sum_{i=0}^{n_{\rm 3b}} \mathrm{P}(i, \kappa_{\rm 3b} R \Delta t_{\rm 3b} C_{\rm 3b}) \frac{1}{\Delta t_{\rm 3b}^{i}} \left[ \int_{0}^{\Delta t_{\rm 3b}} \left[1 - \left(t_{\rm (3b,4a_e)} - t_{\rm (3b,4a_b)} + t_{\rm (3b,3b_e)} \right)\right]d\tau_{\rm 3b} \right]^{i} \right] \\
    & \times \left[ \sum_{i=0}^{n_{\rm 4a}} \mathrm{P}(i, \kappa_{\rm 4a} R \Delta t_{\rm 4a} C_{\rm 4a}) \frac{1}{\Delta t_{\rm 4a}^{i}} \left[ \int_{0}^{\Delta t_{\rm 4a}} \left[1 - t_{\rm (4a,4a_e)} \right]d\tau_{\rm 4a} \right]^{i} \right] \\
    & \times \left[ \int_{0}^{\Delta t_{\rm 3a}} \left[t_{\rm (3a,4a_e)} - t_{\rm (3a,4a_b)} + t_{\rm (3a,3b_e)} - t_{\rm (3a,3b_b)} + t_{\rm (3a,3a_e)}\right]d\tau_{\rm 3a} \right]
\mbox{,}\end{split}
\end{equation}
\begin{equation}
\begin{split}
    & P_{\rm 3b} = \\
    & \left[ \sum_{i=0}^{n_1} \mathrm{P}(i, \kappa_1 R \Delta t_1 C_1) \frac{1}{\Delta t_1^{i}} \left[ \int_{0}^{\Delta t_1} \left[1 - \left(t_{\rm (1,4a_e)} - t_{\rm (1,4a_b)} + t_{\rm (1,3b_e)} - t_{\rm (1,3b_b)} + t_{\rm (1,3a_e)} - t_{\rm (1,3a_b)} + t_{\rm (1,2_e)} - t_{\rm (1,2_b)} + t_{\rm (1,1_e)}\right)\right]d\tau_1 \right]^{i} \right] \\
    & \times \left[ \sum_{i=0}^{n_2} \mathrm{P}(i, \kappa_2 R \Delta t_2 C_2) \frac{1}{\Delta t_2^{i}} \left[ \int_{0}^{\Delta t_2} \left[1 - \left(t_{\rm (2,4a_e)} - t_{\rm (2,4a_b)} + t_{\rm (2,3b_e)} - t_{\rm (2,3b_b)} + t_{\rm (2,3a_e)} - t_{\rm (2,3a_b)} + t_{\rm (2,2_e)} \right)\right]d\tau_2 \right]^{i} \right] \\
    & \times \left[ \sum_{i=0}^{n_{\rm 3a}} \mathrm{P}(i, \kappa_{\rm 3a} R \Delta t_{\rm 3a} C_{\rm 3a}) \frac{1}{\Delta t_{\rm 3a}^{i}} \left[ \int_{0}^{\Delta t_{\rm 3a}} \left[1 - \left(t_{\rm (3a,4a_e)} - t_{\rm (3a,4a_b)} + t_{\rm (3a,3b_e)} - t_{\rm (3a,3b_b)} + t_{\rm (3a,3a_e)} \right)\right]d\tau_{\rm 3a} \right]^{i} \right] \\
    & \times \left[ \sum_{i=1}^{n_{\rm 3b}} \mathrm{P}(i, \kappa_{\rm 3b} R \Delta t_{\rm 3b} C_{\rm 3b}) \frac{i}{\Delta t_{\rm 3b}^{i}} \left[ \int_{0}^{\Delta t_{\rm 3b}} \left[1 - \left(t_{\rm (3b,4a_e)} - t_{\rm (3b,4a_b)} + t_{\rm (3b,3b_e)} \right)\right]d\tau_{\rm 3b} \right]^{i-1} \right] \\
    & \times \left[ \sum_{i=0}^{n_{\rm 4a}} \mathrm{P}(i, \kappa_{\rm 4a} R \Delta t_{\rm 4a} C_{\rm 4a}) \frac{1}{\Delta t_{\rm 4a}^{i}} \left[ \int_{0}^{\Delta t_{\rm 4a}} \left[1 - t_{\rm (4a,4a_e)} \right]d\tau_{\rm 4a} \right]^{i} \right] \\
    & \times \left[ \int_{0}^{\Delta t_{\rm 3b}} \left[t_{\rm (3b,4a_e)} - t_{\rm (3b,4a_b)} + t_{\rm (3b,3b_e)}\right]d\tau_{\rm 3b} \right]
\mbox{,}\end{split}
\end{equation}
and
\begin{equation}
\begin{split}
    & P_{\rm 4a} = \\
    & \left[ \sum_{i=0}^{n_1} \mathrm{P}(i, \kappa_1 R \Delta t_1 C_1) \frac{1}{\Delta t_1^{i}} \left[ \int_{0}^{\Delta t_1} \left[1 - \left(t_{\rm (1,4a_e)} - t_{\rm (1,4a_b)} + t_{\rm (1,3b_e)} - t_{\rm (1,3b_b)} + t_{\rm (1,3a_e)} - t_{\rm (1,3a_b)} + t_{\rm (1,2_e)} - t_{\rm (1,2_b)} + t_{\rm (1,1_e)}\right)\right]d\tau_1 \right]^{i} \right] \\
    & \times \left[ \sum_{i=0}^{n_2} \mathrm{P}(i, \kappa_2 R \Delta t_2 C_2) \frac{1}{\Delta t_2^{i}} \left[ \int_{0}^{\Delta t_2} \left[1 - \left(t_{\rm (2,4a_e)} - t_{\rm (2,4a_b)} + t_{\rm (2,3b_e)} - t_{\rm (2,3b_b)} + t_{\rm (2,3a_e)} - t_{\rm (2,3a_b)} + t_{\rm (2,2_e)} \right)\right]d\tau_2 \right]^{i} \right] \\
    & \times \left[ \sum_{i=0}^{n_{\rm 3a}} \mathrm{P}(i, \kappa_{\rm 3a} R \Delta t_{\rm 3a} C_{\rm 3a}) \frac{1}{\Delta t_{\rm 3a}^{i}} \left[ \int_{0}^{\Delta t_{\rm 3a}} \left[1 - \left(t_{\rm (3a,4a_e)} - t_{\rm (3a,4a_b)} + t_{\rm (3a,3b_e)} - t_{\rm (3a,3b_b)} + t_{\rm (3a,3a_e)} \right)\right]d\tau_{\rm 3a} \right]^{i} \right] \\
    & \times \left[ \sum_{i=0}^{n_{\rm 3b}} \mathrm{P}(i, \kappa_{\rm 3b} R \Delta t_{\rm 3b} C_{\rm 3b}) \frac{1}{\Delta t_{\rm 3b}^{i}} \left[ \int_{0}^{\Delta t_{\rm 3b}} \left[1 - \left(t_{\rm (3b,4a_e)} - t_{\rm (3b,4a_b)} + t_{\rm (3b,3b_e)} \right)\right]d\tau_{\rm 3b} \right]^{i} \right] \\
    & \times \left[ \sum_{i=1}^{n_{\rm 4a}} \mathrm{P}(i, \kappa_{\rm 4a} R \Delta t_{\rm 4a} C_{\rm 4a}) \frac{i}{\Delta t_{\rm 4a}^{i}} \left[ \int_{0}^{\Delta t_{\rm 4a}} \left[1 - t_{\rm (4a,4a_e)} \right]d\tau_{\rm 4a} \right]^{i-1} \right] \\
    & \times \left[ \int_{0}^{\Delta t_{\rm 4a}} t_{\rm (4a,4a_e)} d\tau_{\rm 4a} \right]
\mbox{.}\end{split}
\end{equation}

\begin{acknowledgments}

We are grateful to the anonymous referee for the insightful suggestions. We thank Yin-Jie Li for the valuable discussions and sharing the population posterior data. We also thank Do{\u{g}}a Veske for the helpful correspondence. This work is supported in part by the National Key R\&D Program of China (Grants No. 2021YFA0718500), the National Natural Science Foundation of China under Grants No. 12603087, No. 12588101, No. 12233011, No. 12303056, No. 12233002, and No. 12473042. This research has made use of data and software obtained from the Gravitational Wave Open Science Center (\url{https://www.gw-openscience.org}), a service of LIGO Laboratory, the LIGO Scientific Collaboration, the Virgo Collaboration and KAGRA. LIGO Laboratory and Advanced LIGO are funded by the United States National Science Foundation (NSF) as well as the Science and Technology Facilities Council (STFC) of the United Kingdom, the Max-Planck-Society (MPS), the State of Niedersachsen/Germany, and the Australian Research Council. Virgo is funded, through the European Gravitationa Observatory (EGO), by the French Center National de Recherche Scientifique (CNRS), the Italian Istituto Nazionale di Fisica Nucleare (INFN), and the Dutch Nikhef, with contributions by institutions from Belgium, Germany, Greece, Hungary, Ireland, Japan, Monaco, Poland, Portugal, Spain. KAGRA are funded by Ministry of Education, Culture, Sports, Science and Technology (MEXT), and Japan Society for the Promotion of Science (JSPS), National Research Foundation (NRF) and Ministry of Science and ICT (MSIT) in Korea, Academia Sinica (AS) and the Ministry of Science and Technology (MoST) in Taiwan.

\end{acknowledgments}

\bibliography{re}

@ARTICLE{2022ApJ...941L..39W,
       author = {{Wang}, Yuan-Zhu and {Li}, Yin-Jie and {Vink}, Jorick S. and {Fan}, Yi-Zhong and {Tang}, Shao-Peng and {Qin}, Ying and {Wei}, Da-Ming},
        title = "{Potential Subpopulations and Assembling Tendency of the Merging Black Holes}",
      journal = {\apjl},
         year = 2022,
        month = dec,
       volume = {941},
       number = {2},
          eid = {L39},
        pages = {L39},
          doi = {10.3847/2041-8213/aca89f},
archivePrefix = {arXiv},
       eprint = {2208.11871},
 primaryClass = {astro-ph.HE},
       adsurl = {https://ui.adsabs.harvard.edu/abs/2022ApJ...941L..39W}
}

@ARTICLE{2025PhRvL.135k1403A,
       author = {{Abac}, A.~G. and {Abouelfettouh}, I. and {Acernese}, F. and {Ackley}, K. and {Adamcewicz}, C. and {Adhicary}, S. and {Adhikari}, D. and {Adhikari}, N. and others},
        title = "{GW250114: Testing Hawking's Area Law and the Kerr Nature of Black Holes}",
      journal = {\prl},
         year = 2025,
        month = sep,
       volume = {135},
       number = {11},
          eid = {111403},
        pages = {111403},
          doi = {10.1103/kw5g-d732},
archivePrefix = {arXiv},
       eprint = {2509.08054},
 primaryClass = {gr-qc},
       adsurl = {https://ui.adsabs.harvard.edu/abs/2025PhRvL.135k1403A}
}

@ARTICLE{2026SciBu..71...83T,
       author = {{Tang}, Shao-Peng and {Wang}, Hai-Tian and {Li}, Yin-Jie and {Fan}, Yi-Zhong},
        title = "{Verification of the black hole area law with GW230814}",
      journal = {Sci. Bull.},
         year = 2026,
        month = jan,
       volume = {71},
       number = {1},
        pages = {83-88},
          doi = {10.1016/j.scib.2025.11.002},
archivePrefix = {arXiv},
       eprint = {2509.03480},
 primaryClass = {gr-qc},
       adsurl = {https://ui.adsabs.harvard.edu/abs/2026SciBu..71...83T}
}

@ARTICLE{2026PhRvL.136d1403A,
       author = {{Abac}, A.~G. and {Abouelfettouh}, I. and {Acernese}, F. and {Ackley}, K. and {Adamcewicz}, C. and {Adhicary}, S. and {Adhikari}, D. and {Adhikari}, N. and others},
        title = "{Black Hole Spectroscopy and Tests of General Relativity with GW250114}",
      journal = {\prl},
         year = 2026,
        month = jan,
       volume = {136},
       number = {4},
          eid = {041403},
        pages = {041403},
          doi = {10.1103/6c61-fm1n},
archivePrefix = {arXiv},
       eprint = {2509.08099},
 primaryClass = {gr-qc},
       adsurl = {https://ui.adsabs.harvard.edu/abs/2026PhRvL.136d1403A}
}

@ARTICLE{2021arXiv211010838W,
       author = {{Wang}, Yuan-Zhu and {Fan}, Yi-Zhong and {Tang}, Shao-Peng and {Qin}, Ying and {Wei}, Da-Ming},
        title = "{Mass and effective spin distributions of binary black hole events: Evidence for a population of hierarchical black hole mergers in active galactic nuclei disks}",
      journal = {arXiv e-prints},
         year = 2021,
        month = oct,
          eid = {arXiv:2110.10838},
        pages = {arXiv:2110.10838},
          doi = {10.48550/arXiv.2110.10838},
archivePrefix = {arXiv},
       eprint = {2110.10838},
 primaryClass = {astro-ph.HE},
       adsurl = {https://ui.adsabs.harvard.edu/abs/2021arXiv211010838W}
}

@ARTICLE{2002MNRAS.330..232C,
       author = {{Miller}, M. Coleman and {Hamilton}, Douglas P.},
        title = "{Production of intermediate-mass black holes in globular clusters}",
      journal = {\mnras},
         year = 2002,
        month = feb,
       volume = {330},
       number = {1},
        pages = {232-240},
          doi = {10.1046/j.1365-8711.2002.05112.x},
archivePrefix = {arXiv},
       eprint = {astro-ph/0106188},
 primaryClass = {astro-ph},
       adsurl = {https://ui.adsabs.harvard.edu/abs/2002MNRAS.330..232C}
}

@ARTICLE{2026SCPMA..6999562W,
       author = {{Wang}, Yuan-Zhu and {Li}, Yin-Jie and {Gao}, Shi-Jie and {Tang}, Shao-Peng and {Fan}, Yi-Zhong},
        title = "{A new group of low-spin 50M$_{{\ensuremath{\odot}}}$─70M$_{{\ensuremath{\odot}}}$ black holes and the high pair-instability mass cutoff}",
      journal = {Science China Physics, Mechanics, and Astronomy},
         year = 2026,
        month = may,
       volume = {69},
       number = {9},
          eid = {299562},
        pages = {299562},
          doi = {10.1007/s11433-026-3004-6},
archivePrefix = {arXiv},
       eprint = {2510.22698},
 primaryClass = {astro-ph.HE},
       adsurl = {https://ui.adsabs.harvard.edu/abs/2026SCPMA..6999562W}
}

@ARTICLE{2025ApJ...993L..25A,
       author = {{Abac}, A.~G. and {Abouelfettouh}, I. and {Acernese}, F. and {Ackley}, K. and {Adamcewicz}, C. and {Adhicary}, S. and {Adhikari}, D. and {Adhikari}, N. and {Adhikari}, R.~X. and {Adkins}, V.~K. and {Afroz}, S. and {Agapito}, A. and {Agarwal}, D. and {Agathos}, M. and {Aggarwal}, N. and {Aggarwal}, S. and {Aguiar}, O.~D. and {Ahrend}, I.-L. and {Aiello}, L. and {Ain}, A. and {Ajith}, P. and {Akutsu}, T. and {Albanesi}, S. and {Ali}, W. and {Al-Kershi}, S. and {All{\'e}n{\'e}}, C. and {Allocca}, A. and {Al-Shammari}, S. and {Altin}, P.~A. and {Alvarez-Lopez}, S. and {Amar}, W. and {Amarasinghe}, O. and {Amato}, A. and {Amicucci}, F. and {Amra}, C. and {Ananyeva}, A. and {Anderson}, S.~B. and {Anderson}, W.~G. and {Andia}, M. and {Ando}, M. and {Andr{\'e}s-Carcasona}, M. and {Andri{\'c}}, T. and {Anglin}, J. and {Ansoldi}, S. and {Antelis}, J.~M. and {Antier}, S. and {Aoumi}, M. and {Appavuravther}, E.~Z. and {Appert}, S. and {Apple}, S.~K. and {Arai}, K. and {Alvarez}, C. Araujo and {Araya}, A. and {Araya}, M.~C. and {Arca Sedda}, M. and {Areeda}, J.~S. and {Aritomi}, N. and {Armato}, F. and {Armstrong}, S. and {Arnaud}, N. and {Arogeti}, M. and {Aronson}, S.~M. and {Arun}, K.~G. and {Ashton}, G. and {Aso}, Y. and {Asprea}, L. and {Assiduo}, M. and {Assis de Souza Melo}, S. and {Aston}, S.~M. and {Astone}, P. and {Attadio}, F. and {Aubin}, F. and {AultONeal}, K. and {Avallone}, G. and {Avila}, E.~A. and {Babak}, S. and {Badger}, C. and {Bae}, S. and {Bagnasco}, S. and {Baiotti}, L. and {Bajpai}, R. and {Baka}, T. and {Baker}, A.~M. and {Baker}, K.~A. and {Baker}, T. and {Baldi}, G. and {Baldicchi}, N. and {Ball}, M. and {Ballardin}, G. and {Ballmer}, S.~W. and {Banagiri}, S. and {Banerjee}, B. and {Bankar}, D. and {Baptiste}, T.~M. and {Baral}, P. and {Baratti}, M. and {Barayoga}, J.~C. and {Barish}, B.~C. and {Barker}, D. and {Barman}, N. and {Barneo}, P. and {Barone}, F. and {Barr}, B. and {Barsotti}, L. and {Barsuglia}, M. and {Barta}, D. and {Bartoletti}, A.~M. and {Barton}, M.~A. and {Bartos}, I. and {Basalaev}, A. and {Bassiri}, R. and {Basti}, A. and {Bawaj}, M. and {Baxi}, P. and {Bayley}, J.~C. and {Baylor}, A.~C. and {Baynard}, II, P.~A. and {Bazzan}, M. and {Bedakihale}, V.~M. and {Beirnaert}, F. and {Bejger}, M. and {Belardinelli}, D. and {Bell}, A.~S. and {Bellie}, D.~S. and {Bellizzi}, L. and {Benoit}, W. and {Bentara}, I. and {Bentley}, J.~D. and {Ben Yaala}, M. and {Bera}, S. and {Bergamin}, F. and {Berger}, B.~K. and {Bernuzzi}, S. and {Beroiz}, M. and {Berry}, C.~P.~L. and {Bersanetti}, D. and {Bertheas}, T. and {Bertolini}, A. and {Betzwieser}, J. and {Beveridge}, D. and {Bevilacqua}, G. and {Bevins}, N. and {Bhandare}, R. and {Bhatt}, R. and {Bhattacharjee}, D. and {Bhattacharyya}, S. and {Bhaumik}, S. and {Bhagwat}, S. and {Biancalana}, V. and {Bianchi}, A. and {Bilenko}, I.~A. and {Billingsley}, G. and {Binetti}, A. and {Bini}, S. and {Binu}, C. and {Biot}, S. and {Birnholtz}, O. and {Biscoveanu}, S. and {Bisht}, A. and {Bitossi}, M. and {Bizouard}, M.-A. and {Blaber}, S. and {Blackburn}, J.~K. and {Blagg}, L.~A. and {Blair}, C.~D. and {Blair}, D.~G. and {Bode}, N. and {Boettner}, N. and {Boileau}, G. and {Boldrini}, M. and {Bolingbroke}, G.~N. and {Bolliand}, A. and {Bonavena}, L.~D. and {Bondarescu}, R. and {Bondu}, F. and {Bonilla}, E. and {Bonilla}, M.~S. and {Bonino}, A. and {Bonnand}, R. and {Borchers}, A. and {Borhanian}, S. and {Boschi}, V. and {Bose}, S. and {Bossilkov}, V. and {Bothra}, Y. and {Boudon}, A. and {Bourg}, L. and {Boyle}, M. and {Bozzi}, A. and {Bradaschia}, C. and {Brady}, P.~R. and {Branch}, A. and {Branchesi}, M. and {Braun}, I. and {Briant}, T. and {Brillet}, A. and {Brinkmann}, M. and {Brockill}, P. and {Brockmueller}, E. and {Brooks}, A.~F.},
        title = "{GW231123: A Binary Black Hole Merger with Total Mass 190─265 M$_{{\ensuremath{\odot}}}$}",
      journal = {\apjl},
         year = 2025,
        month = nov,
       volume = {993},
       number = {1},
          eid = {L25},
        pages = {L25},
          doi = {10.3847/2041-8213/ae0c9c},
archivePrefix = {arXiv},
       eprint = {2507.08219},
 primaryClass = {astro-ph.HE},
       adsurl = {https://ui.adsabs.harvard.edu/abs/2025ApJ...993L..25A}
}

@ARTICLE{2018MNRAS.480L..58A,
       author = {{Antonini}, Fabio and {Rodriguez}, Carl L. and {Petrovich}, Cristobal and {Fischer}, Caitlin L.},
        title = "{Precessional dynamics of black hole triples: binary mergers with near-zero effective spin}",
      journal = {\mnras},
         year = 2018,
        month = oct,
       volume = {480},
       number = {1},
        pages = {L58-L62},
          doi = {10.1093/mnrasl/sly126},
archivePrefix = {arXiv},
       eprint = {1711.07142},
 primaryClass = {astro-ph.HE},
       adsurl = {https://ui.adsabs.harvard.edu/abs/2018MNRAS.480L..58A}
}

@ARTICLE{2018ApJ...863...68L,
       author = {{Liu}, Bin and {Lai}, Dong},
        title = "{Black Hole and Neutron Star Binary Mergers in Triple Systems: Merger Fraction and Spin-Orbit Misalignment}",
      journal = {\apj},
         year = 2018,
        month = aug,
       volume = {863},
       number = {1},
          eid = {68},
        pages = {68},
          doi = {10.3847/1538-4357/aad09f},
archivePrefix = {arXiv},
       eprint = {1805.03202},
 primaryClass = {astro-ph.HE},
       adsurl = {https://ui.adsabs.harvard.edu/abs/2018ApJ...863...68L}
}

@ARTICLE{2021NatAs...5..749G,
       author = {{Gerosa}, Davide and {Fishbach}, Maya},
        title = "{Hierarchical mergers of stellar-mass black holes and their gravitational-wave signatures}",
      journal = {Nature Astronomy},
         year = 2021,
        month = jul,
       volume = {5},
        pages = {749-760},
          doi = {10.1038/s41550-021-01398-w},
archivePrefix = {arXiv},
       eprint = {2105.03439},
 primaryClass = {astro-ph.HE},
       adsurl = {https://ui.adsabs.harvard.edu/abs/2021NatAs...5..749G}
}

@ARTICLE{2025ApJ...987...65L,
       author = {{Li}, Yin-Jie and {Wang}, Yuan-Zhu and {Tang}, Shao-Peng and {Chen}, Tong and {Fan}, Yi-Zhong},
        title = "{Revealing the {\ensuremath{\chi}}$_{eff}$─q Correlation among Coalescing Binary Black Holes and Tentative Evidence for AGN-driven Hierarchical Mergers}",
      journal = {\apj},
         year = 2025,
        month = jul,
       volume = {987},
       number = {1},
          eid = {65},
        pages = {65},
          doi = {10.3847/1538-4357/add535},
archivePrefix = {arXiv},
       eprint = {2501.09495},
 primaryClass = {astro-ph.HE},
       adsurl = {https://ui.adsabs.harvard.edu/abs/2025ApJ...987...65L}
}

@ARTICLE{2023PhRvD.108l4037R,
       author = {{Ramos-Buades}, Antoni and {Buonanno}, Alessandra and {Estell{\'e}s}, H{\'e}ctor and {Khalil}, Mohammed and {Mihaylov}, Deyan P. and {Ossokine}, Serguei and {Pompili}, Lorenzo and {Shiferaw}, Mahlet},
        title = "{Next generation of accurate and efficient multipolar precessing-spin effective-one-body waveforms for binary black holes}",
      journal = {\prd},
         year = 2023,
        month = dec,
       volume = {108},
       number = {12},
          eid = {124037},
        pages = {124037},
          doi = {10.1103/PhysRevD.108.124037},
archivePrefix = {arXiv},
       eprint = {2303.18046},
 primaryClass = {gr-qc},
       adsurl = {https://ui.adsabs.harvard.edu/abs/2023PhRvD.108l4037R}
}

@ARTICLE{2019ApJ...881...41L,
       author = {{Liu}, Bin and {Lai}, Dong and {Wang}, Yi-Han},
        title = "{Black Hole and Neutron Star Binary Mergers in Triple Systems. II. Merger Eccentricity and Spin-Orbit Misalignment}",
      journal = {\apj},
         year = 2019,
        month = aug,
       volume = {881},
       number = {1},
          eid = {41},
        pages = {41},
          doi = {10.3847/1538-4357/ab2dfb},
archivePrefix = {arXiv},
       eprint = {1905.00427},
 primaryClass = {astro-ph.HE},
       adsurl = {https://ui.adsabs.harvard.edu/abs/2019ApJ...881...41L}
}

@ARTICLE{2001MNRAS.321..398M,
       author = {{Mardling}, Rosemary A. and {Aarseth}, Sverre J.},
        title = "{Tidal interactions in star cluster simulations}",
      journal = {\mnras},
         year = 2001,
        month = mar,
       volume = {321},
       number = {3},
        pages = {398-420},
          doi = {10.1046/j.1365-8711.2001.03974.x},
       adsurl = {https://ui.adsabs.harvard.edu/abs/2001MNRAS.321..398M}
}

@ARTICLE{2010PhRvD..82j7501B,
       author = {{Berry}, Christopher P.~L. and {Gair}, Jonathan R.},
        title = "{Gravitational wave energy spectrum of a parabolic encounter}",
      journal = {\prd},
         year = 2010,
        month = nov,
       volume = {82},
       number = {10},
          eid = {107501},
        pages = {107501},
          doi = {10.1103/PhysRevD.82.107501},
archivePrefix = {arXiv},
       eprint = {1010.3865},
 primaryClass = {gr-qc},
       adsurl = {https://ui.adsabs.harvard.edu/abs/2010PhRvD..82j7501B}
}

@ARTICLE{2012PhRvD..86d4017D,
       author = {{De Vittori}, Lorenzo and {Jetzer}, Philippe and {Klein}, Antoine},
        title = "{Gravitational wave energy spectrum of hyperbolic encounters}",
      journal = {\prd},
         year = 2012,
        month = aug,
       volume = {86},
       number = {4},
          eid = {044017},
        pages = {044017},
          doi = {10.1103/PhysRevD.86.044017},
archivePrefix = {arXiv},
       eprint = {1207.5359},
 primaryClass = {gr-qc},
       adsurl = {https://ui.adsabs.harvard.edu/abs/2012PhRvD..86d4017D}
}

@ARTICLE{2018MNRAS.481.5436S,
       author = {{Samsing}, Johan and {Leigh}, Nathan W.~C. and {Trani}, Alessandro A.},
        title = "{Implementing tidal and gravitational wave energy losses in few-body codes: A fast and easy drag force model}",
      journal = {\mnras},
         year = 2018,
        month = dec,
       volume = {481},
       number = {4},
        pages = {5436-5444},
          doi = {10.1093/mnras/sty2247},
archivePrefix = {arXiv},
       eprint = {1803.08215},
 primaryClass = {astro-ph.HE},
       adsurl = {https://ui.adsabs.harvard.edu/abs/2018MNRAS.481.5436S}
}

@ARTICLE{1963PhRv..131..435P,
       author = {{Peters}, P.~C. and {Mathews}, J.},
        title = "{Gravitational Radiation from Point Masses in a Keplerian Orbit}",
      journal = {Physical Review},
         year = 1963,
        month = jul,
       volume = {131},
       number = {1},
        pages = {435-440},
          doi = {10.1103/PhysRev.131.435},
       adsurl = {https://ui.adsabs.harvard.edu/abs/1963PhRv..131..435P}
}

@ARTICLE{1985AIHPA..43..107D,
       author = {{Damour}, T. and {Deruelle}, N.},
        title = "{General relativistic celestial mechanics of binary systems. I. The post-Newtonian motion.}",
      journal = {Annales de L'Institut Henri Poincare Section (A) Physique Theorique},
         year = 1985,
        month = jan,
       volume = {43},
       number = {1},
        pages = {107-132},
       adsurl = {https://ui.adsabs.harvard.edu/abs/1985AIHPA..43..107D}
}

@BOOK{1989racm.book.....S,
       author = {{Soffel}, Michael H.},
        title = "{Relativity in Astrometry, Celestial Mechanics and Geodesy}",
         year = 1989,
    publisher = {Springer-Verlag},
       adsurl = {https://ui.adsabs.harvard.edu/abs/1989racm.book.....S}
}

@ARTICLE{2017ApJ...846...36S,
       author = {{Samsing}, Johan and {MacLeod}, Morgan and {Ramirez-Ruiz}, Enrico},
        title = "{Formation of Tidal Captures and Gravitational Wave Inspirals in Binary-single Interactions}",
      journal = {\apj},
         year = 2017,
        month = sep,
       volume = {846},
       number = {1},
          eid = {36},
        pages = {36},
          doi = {10.3847/1538-4357/aa7e32},
archivePrefix = {arXiv},
       eprint = {1609.09114},
 primaryClass = {astro-ph.HE},
       adsurl = {https://ui.adsabs.harvard.edu/abs/2017ApJ...846...36S}
}

@ARTICLE{2026MNRAS.549ag944G,
       author = {{Ginat}, Yonadav Barry and {Stegmann}, Jakob and {Samsing}, Johan},
        title = "{Dynamical evolution of quasi-hierarchical triples}",
      journal = {\mnras},
         year = 2026,
        month = jun,
       volume = {549},
       number = {2},
          eid = {stag944},
        pages = {stag944},
          doi = {10.1093/mnras/stag944},
archivePrefix = {arXiv},
       eprint = {2509.02685},
 primaryClass = {astro-ph.HE},
       adsurl = {https://ui.adsabs.harvard.edu/abs/2026MNRAS.549ag944G}
}

@ARTICLE{2019MNRAS.482...30S,
       author = {{Samsing}, Johan and {Ilan}, Teva},
        title = "{Double gravitational wave mergers}",
      journal = {\mnras},
         year = 2019,
        month = jan,
       volume = {482},
       number = {1},
        pages = {30-39},
          doi = {10.1093/mnras/sty2249},
archivePrefix = {arXiv},
       eprint = {1709.01660},
 primaryClass = {astro-ph.HE},
       adsurl = {https://ui.adsabs.harvard.edu/abs/2019MNRAS.482...30S}
}

@ARTICLE{1983ApJ...268..319H,
       author = {{Hut}, P. and {Bahcall}, J.~N.},
        title = "{Binary-single star scattering. I - Numerical experiments for equal masses}",
      journal = {\apj},
         year = 1983,
        month = may,
       volume = {268},
        pages = {319-341},
          doi = {10.1086/160956},
       adsurl = {https://ui.adsabs.harvard.edu/abs/1983ApJ...268..319H}
}

@ARTICLE{1993ApJ...403..256H,
       author = {{Hut}, Piet},
        title = "{Binary--Single-Star Scattering. III. Numerical Experiments for Equal-Mass Hard Binaries}",
      journal = {\apj},
         year = 1993,
        month = jan,
       volume = {403},
        pages = {256},
          doi = {10.1086/172199},
       adsurl = {https://ui.adsabs.harvard.edu/abs/1993ApJ...403..256H}
}

@ARTICLE{2013MNRAS.431.2155N,
       author = {{Naoz}, Smadar and {Farr}, Will M. and {Lithwick}, Yoram and {Rasio}, Frederic A. and {Teyssandier}, Jean},
        title = "{Secular dynamics in hierarchical three-body systems}",
      journal = {\mnras},
         year = 2013,
        month = may,
       volume = {431},
       number = {3},
        pages = {2155-2171},
          doi = {10.1093/mnras/stt302},
archivePrefix = {arXiv},
       eprint = {1107.2414},
 primaryClass = {astro-ph.EP},
       adsurl = {https://ui.adsabs.harvard.edu/abs/2013MNRAS.431.2155N}
}

@ARTICLE{2014ApJ...784...71S,
       author = {{Samsing}, Johan and {MacLeod}, Morgan and {Ramirez-Ruiz}, Enrico},
        title = "{The Formation of Eccentric Compact Binary Inspirals and the Role of Gravitational Wave Emission in Binary-Single Stellar Encounters}",
      journal = {\apj},
         year = 2014,
        month = mar,
       volume = {784},
       number = {1},
          eid = {71},
        pages = {71},
          doi = {10.1088/0004-637X/784/1/71},
archivePrefix = {arXiv},
       eprint = {1308.2964},
 primaryClass = {astro-ph.HE},
       adsurl = {https://ui.adsabs.harvard.edu/abs/2014ApJ...784...71S}
}

@ARTICLE{2025A&A...697A.118R,
       author = {{Rando Forastier}, Bruno and {Mar{\'\i}n Pina}, Daniel and {Gieles}, Mark and {Portegies Zwart}, Simon and {Antonini}, Fabio},
        title = "{Binary-single interactions with different mass ratios: Implications for gravitational waves from globular clusters}",
      journal = {\aap},
         year = 2025,
        month = may,
       volume = {697},
          eid = {A118},
        pages = {A118},
          doi = {10.1051/0004-6361/202450890},
archivePrefix = {arXiv},
       eprint = {2405.16999},
 primaryClass = {astro-ph.GA},
       adsurl = {https://ui.adsabs.harvard.edu/abs/2025A&A...697A.118R}
}

@article{JeanChazy1929,
author = {Jean Chazy},
journal = {Journal de Mathématiques Pures et Appliquées},
language = {fre},
pages = {353-380},
title = {Sur l'allure finale du mouvement dans le problème des trois corps},
url = {http://eudml.org/doc/234704},
volume = {8},
year = {1929},
}

@ARTICLE{1975MNRAS.173..729H,
       author = {{Heggie}, D.~C.},
        title = "{Binary evolution in stellar dynamics.}",
      journal = {\mnras},
         year = 1975,
        month = dec,
       volume = {173},
        pages = {729-787},
          doi = {10.1093/mnras/173.3.729},
       adsurl = {https://ui.adsabs.harvard.edu/abs/1975MNRAS.173..729H}
}

@ARTICLE{2019ApJ...871...91Z,
       author = {{Zevin}, Michael and {Samsing}, Johan and {Rodriguez}, Carl and {Haster}, Carl-Johan and {Ramirez-Ruiz}, Enrico},
        title = "{Eccentric Black Hole Mergers in Dense Star Clusters: The Role of Binary-Binary Encounters}",
      journal = {\apj},
         year = 2019,
        month = jan,
       volume = {871},
       number = {1},
          eid = {91},
        pages = {91},
          doi = {10.3847/1538-4357/aaf6ec},
archivePrefix = {arXiv},
       eprint = {1810.00901},
 primaryClass = {astro-ph.HE},
       adsurl = {https://ui.adsabs.harvard.edu/abs/2019ApJ...871...91Z}
}

@ARTICLE{2016MNRAS.456.4219A,
       author = {{Antognini}, Joseph M.~O. and {Thompson}, Todd A.},
        title = "{Dynamical formation and scattering of hierarchical triples: cross-sections, Kozai-Lidov oscillations, and collisions}",
      journal = {\mnras},
         year = 2016,
        month = mar,
       volume = {456},
       number = {4},
        pages = {4219-4246},
          doi = {10.1093/mnras/stv2938},
archivePrefix = {arXiv},
       eprint = {1507.03593},
 primaryClass = {astro-ph.SR},
       adsurl = {https://ui.adsabs.harvard.edu/abs/2016MNRAS.456.4219A}
}

@ARTICLE{2025PASP..137i4201S,
       author = {{Shariat}, Cheyanne and {El-Badry}, Kareem and {Naoz}, Smadar},
        title = "{10,000 Resolved Triples from Gaia: Empirical Constraints on Triple Star Populations}",
      journal = {\pasp},
         year = 2025,
        month = sep,
       volume = {137},
       number = {9},
          eid = {094201},
        pages = {094201},
          doi = {10.1088/1538-3873/adfb30},
archivePrefix = {arXiv},
       eprint = {2506.16513},
 primaryClass = {astro-ph.SR},
       adsurl = {https://ui.adsabs.harvard.edu/abs/2025PASP..137i4201S}
}

@ARTICLE{2018ApJ...863....7R,
       author = {{Rodriguez}, Carl L. and {Antonini}, Fabio},
        title = "{A Triple Origin for the Heavy and Low-spin Binary Black Holes Detected by LIGO/VIRGO}",
      journal = {\apj},
         year = 2018,
        month = aug,
       volume = {863},
       number = {1},
          eid = {7},
        pages = {7},
          doi = {10.3847/1538-4357/aacea4},
archivePrefix = {arXiv},
       eprint = {1805.08212},
 primaryClass = {astro-ph.HE},
       adsurl = {https://ui.adsabs.harvard.edu/abs/2018ApJ...863....7R}
}

@ARTICLE{2026PhRvL.137b1404P,
       author = {{Plunkett}, Cailin and {Vitale}, Salvatore and {Callister}, Thomas and {Zevin}, Michael and {(Society Of Physicists Interested in Non-Aligned Spins (Spins))}},
        title = "{Signatures of a Subpopulation of Hierarchical Mergers in the GWTC-4 Gravitational-Wave Dataset}",
      journal = {\prl},
         year = 2026,
        month = jul,
       volume = {137},
       number = {2},
          eid = {021404},
        pages = {021404},
          doi = {10.1103/n6p4-ftgq},
archivePrefix = {arXiv},
       eprint = {2601.07908},
 primaryClass = {gr-qc},
       adsurl = {https://ui.adsabs.harvard.edu/abs/2026PhRvL.137b1404P}
}

@article{Arnaud:2026hzn,
    author = "Arnaud, Nicolas and Driggers, Jenne and O{\textquoteright}Reilly, Brian and Sawada, Takahiro",
    collaboration = "LIGO Scientific, VIRGO, KAGRA",
    title = "{The LIGO-Virgo-KAGRA Observing Run 4}",
    doi = "10.1088/1742-6596/3177/1/012067",
    journal = "J. Phys. Conf. Ser.",
    volume = "3177",
    number = "1",
    pages = "012067",
    year = "2026"
}

@ARTICLE{2025arXiv250923897L,
       author = {{Li}, Yin-Jie and {Wang}, Yuan-Zhu and {Tang}, Shao-Peng and {Fan}, Yi-Zhong},
        title = "{Aligned Hierarchical Black Hole Mergers in Active-Galactic-Nuclei Disks Revealed by GWTC-4}",
      journal = {\prl},
         year = 2026,
        month = jul,
       volume = {137},
       number = {2},
          eid = {021407},
        pages = {021407},
          doi = {10.1103/qyfv-wkv1},
archivePrefix = {arXiv},
       eprint = {2509.23897},
 primaryClass = {astro-ph.HE},
       adsurl = {https://ui.adsabs.harvard.edu/abs/2026PhRvL.137b1407L}
}

@ARTICLE{2025MNRAS.539.1501R,
       author = {{Rowan}, Connar and {Whitehead}, Henry and {Fabj}, Gaia and {Saini}, Pankaj and {Kocsis}, Bence and {Pessah}, Martin and {Samsing}, Johan},
        title = "{Prompt gravitational-wave mergers aided by gas in active galactic nuclei: the hydrodynamics of binary-single black hole scatterings}",
      journal = {\mnras},
         year = 2025,
        month = may,
       volume = {539},
       number = {2},
        pages = {1501-1515},
          doi = {10.1093/mnras/staf547},
archivePrefix = {arXiv},
       eprint = {2501.09017},
 primaryClass = {astro-ph.GA},
       adsurl = {https://ui.adsabs.harvard.edu/abs/2025MNRAS.539.1501R}
}

@ARTICLE{2025ApJ...983..114W,
       author = {{Wang}, Mengye and {Ma}, Yiqiu and {Li}, Hui and {Wu}, Qingwen and {Li}, Ya-Ping and {Lei}, Xiangli and {Wu}, Jiancheng},
        title = "{Simulation of Binary-single Interactions in AGN Disk. I. Gas-enhanced Binary Orbital Hardening}",
      journal = {\apj},
         year = 2025,
        month = apr,
       volume = {983},
       number = {2},
          eid = {114},
        pages = {114},
          doi = {10.3847/1538-4357/adbf8e},
archivePrefix = {arXiv},
       eprint = {2501.10703},
 primaryClass = {astro-ph.HE},
       adsurl = {https://ui.adsabs.harvard.edu/abs/2025ApJ...983..114W}
}

@ARTICLE{2025ApJ...993..139W,
       author = {{Wang}, Mengye and {Wu}, Qingwen and {Ma}, Yiqiu},
        title = "{Simulation of Binary─Single Interactions in Active Galactic Nucleus Disks. II. Merger Probability of Binary Black Holes during Chaotic Triple Process}",
      journal = {\apj},
         year = 2025,
        month = nov,
       volume = {993},
       number = {1},
          eid = {139},
        pages = {139},
          doi = {10.3847/1538-4357/ae06a5},
archivePrefix = {arXiv},
       eprint = {2507.07715},
 primaryClass = {astro-ph.HE},
       adsurl = {https://ui.adsabs.harvard.edu/abs/2025ApJ...993..139W}
}

@ARTICLE{2022Natur.603..237S,
       author = {{Samsing}, J. and {Bartos}, I. and {D'Orazio}, D.~J. and {Haiman}, Z. and {Kocsis}, B. and {Leigh}, N.~W.~C. and {Liu}, B. and {Pessah}, M.~E. and {Tagawa}, H.},
        title = "{AGN as potential factories for eccentric black hole mergers}",
      journal = {\nat},
         year = 2022,
        month = mar,
       volume = {603},
       number = {7900},
        pages = {237-240},
          doi = {10.1038/s41586-021-04333-1},
archivePrefix = {arXiv},
       eprint = {2010.09765},
 primaryClass = {astro-ph.HE},
       adsurl = {https://ui.adsabs.harvard.edu/abs/2022Natur.603..237S}
}

@ARTICLE{2023ApJ...954..105V,
       author = {{Vijaykumar}, Aditya and {Tiwari}, Avinash and {Kapadia}, Shasvath J. and {Arun}, K.~G. and {Ajith}, Parameswaran},
        title = "{Waltzing Binaries: Probing the Line-of-sight Acceleration of Merging Compact Objects with Gravitational Waves}",
      journal = {\apj},
         year = 2023,
        month = sep,
       volume = {954},
       number = {1},
          eid = {105},
        pages = {105},
          doi = {10.3847/1538-4357/acd77d},
archivePrefix = {arXiv},
       eprint = {2302.09651},
 primaryClass = {astro-ph.HE},
       adsurl = {https://ui.adsabs.harvard.edu/abs/2023ApJ...954..105V}
}

@ARTICLE{2026PhRvD.113j4017I,
       author = {{Islam}, Tousif and {Wadekar}, Digvijay},
        title = "{Accurate models for recoil velocity distribution in black hole mergers with comparable to extreme mass-ratios and their astrophysical implications}",
      journal = {\prd},
         year = 2026,
        month = may,
       volume = {113},
       number = {10},
          eid = {104017},
        pages = {104017},
          doi = {10.1103/4jvv-qg4h},
archivePrefix = {arXiv},
       eprint = {2511.11536},
 primaryClass = {gr-qc},
       adsurl = {https://ui.adsabs.harvard.edu/abs/2026PhRvD.113j4017I}
}

@ARTICLE{2026arXiv260404546I,
       author = {{Islam}, Tousif},
        title = "{Recoil kicks from binary black hole mergers in GWTC catalogs: implications for retention and hierarchical mergers}",
      journal = {arXiv e-prints},
         year = 2026,
        month = apr,
          eid = {arXiv:2604.04546},
        pages = {arXiv:2604.04546},
          doi = {10.48550/arXiv.2604.04546},
archivePrefix = {arXiv},
       eprint = {2604.04546},
 primaryClass = {astro-ph.HE},
       adsurl = {https://ui.adsabs.harvard.edu/abs/2026arXiv260404546I}
}

@ARTICLE{2002ApJ...568L..23G,
       author = {{Gnedin}, Oleg Y. and {Zhao}, HongSheng and {Pringle}, J.~E. and {Fall}, S. Michael and {Livio}, Mario and {Meylan}, Georges},
        title = "{The Unique History of the Globular Cluster {\ensuremath{\omega}} Centauri}",
      journal = {\apjl},
         year = 2002,
        month = mar,
       volume = {568},
       number = {1},
        pages = {L23-L26},
          doi = {10.1086/340319},
archivePrefix = {arXiv},
       eprint = {astro-ph/0202045},
 primaryClass = {astro-ph},
       adsurl = {https://ui.adsabs.harvard.edu/abs/2002ApJ...568L..23G}
}

@ARTICLE{2019PhRvL.123r1101Y,
       author = {{Yang}, Y. and {Bartos}, I. and {Gayathri}, V. and {Ford}, K.~E.~S. and {Haiman}, Z. and {Klimenko}, S. and {Kocsis}, B. and {M{\'a}rka}, S. and {M{\'a}rka}, Z. and {McKernan}, B. and {O'Shaughnessy}, R.},
        title = "{Hierarchical Black Hole Mergers in Active Galactic Nuclei}",
      journal = {\prl},
         year = 2019,
        month = nov,
       volume = {123},
       number = {18},
          eid = {181101},
        pages = {181101},
          doi = {10.1103/PhysRevLett.123.181101},
archivePrefix = {arXiv},
       eprint = {1906.09281},
 primaryClass = {astro-ph.HE},
       adsurl = {https://ui.adsabs.harvard.edu/abs/2019PhRvL.123r1101Y}
}

@ARTICLE{2024ApJ...965...80G,
       author = {{Gao}, Bo and {Tang}, Shao-Peng and {Yan}, Jingzhi and {Fan}, Yi-Zhong},
        title = "{Prospects of Identifying Hierarchical Triple Mergers for the Third-generation Ground-based Detectors}",
      journal = {\apj},
         year = 2024,
        month = apr,
       volume = {965},
       number = {1},
          eid = {80},
        pages = {80},
          doi = {10.3847/1538-4357/ad2e9a},
archivePrefix = {arXiv},
       eprint = {2402.17381},
 primaryClass = {gr-qc},
       adsurl = {https://ui.adsabs.harvard.edu/abs/2024ApJ...965...80G}
}

@ARTICLE{2023MNRAS.523.4113T,
       author = {{Tang}, Shao-Peng and {Fan}, Yi-Zhong and {Wei}, Da-Ming},
        title = "{Hierarchical triple mergers: testing Hawking's area theorem with the inspiral signals}",
      journal = {\mnras},
         year = 2023,
        month = aug,
       volume = {523},
       number = {3},
        pages = {4113-4119},
          doi = {10.1093/mnras/stad1676},
archivePrefix = {arXiv},
       eprint = {2209.03631},
 primaryClass = {gr-qc},
       adsurl = {https://ui.adsabs.harvard.edu/abs/2023MNRAS.523.4113T}
}

@ARTICLE{2015MNRAS.447..747L,
       author = {{Liu}, Bin and {Mu{\~n}oz}, Diego J. and {Lai}, Dong},
        title = "{Suppression of extreme orbital evolution in triple systems with short-range forces}",
      journal = {\mnras},
         year = 2015,
        month = feb,
       volume = {447},
       number = {1},
        pages = {747-764},
          doi = {10.1093/mnras/stu2396},
archivePrefix = {arXiv},
       eprint = {1409.6717},
 primaryClass = {astro-ph.EP},
       adsurl = {https://ui.adsabs.harvard.edu/abs/2015MNRAS.447..747L}
}

@ARTICLE{2004MNRAS.352....1F,
       author = {{Fregeau}, J.~M. and {Cheung}, P. and {Portegies Zwart}, S.~F. and {Rasio}, F.~A.},
        title = "{Stellar collisions during binary-binary and binary-single star interactions}",
      journal = {\mnras},
         year = 2004,
        month = jul,
       volume = {352},
       number = {1},
        pages = {1-19},
          doi = {10.1111/j.1365-2966.2004.07914.x},
archivePrefix = {arXiv},
       eprint = {astro-ph/0401004},
 primaryClass = {astro-ph},
       adsurl = {https://ui.adsabs.harvard.edu/abs/2004MNRAS.352....1F}
}

@INPROCEEDINGS{2023IAUS..362..404T,
       author = {{Trani}, Alessandro A. and {Spera}, Mario},
        title = "{Modeling gravitational few-body problems with tsunami and okinami}",
    booktitle = {The Predictive Power of Computational Astrophysics as a Discover Tool},
         year = 2023,
       editor = {{Bisikalo}, Dmitry and {Wiebe}, Dmitri and {Boily}, Christian},
       series = {IAU Symposium},
       volume = {362},
        month = jan,
        pages = {404-409},
          doi = {10.1017/S1743921322001818},
archivePrefix = {arXiv},
       eprint = {2206.10583},
 primaryClass = {astro-ph.HE},
       adsurl = {https://ui.adsabs.harvard.edu/abs/2023IAUS..362..404T}
}

@ARTICLE{2012A&A...537A.128R,
       author = {{Rein}, H. and {Liu}, S.-F.},
        title = "{REBOUND: an open-source multi-purpose N-body code for collisional dynamics}",
      journal = {\aap},
         year = 2012,
        month = jan,
       volume = {537},
          eid = {A128},
        pages = {A128},
          doi = {10.1051/0004-6361/201118085},
archivePrefix = {arXiv},
       eprint = {1110.4876},
 primaryClass = {astro-ph.EP},
       adsurl = {https://ui.adsabs.harvard.edu/abs/2012A&A...537A.128R}
}

@article{SpaceHub2021,
title={SpaceHub: A high-performance gravity integration toolkit for few-body problems in astrophysics},
author={Wang, Yihan and Leigh, Nathan and Liu, Bin and Perna, Rosalba},
journal={Monthly Notices of the Royal Astronomical Society},
volume={505},
number={1},
pages={1053--1070},
year={2021},
doi={10.1093/mnras/stab1189}
}

@ARTICLE{1989ApJ...341..175B,
       author = {{Bailyn}, Charles D.},
        title = "{The Formation of Hierarchical Triple Systems by Single-Star--Binary Encounters}",
      journal = {\apj},
         year = 1989,
        month = jun,
       volume = {341},
        pages = {175},
          doi = {10.1086/167482},
       adsurl = {https://ui.adsabs.harvard.edu/abs/1989ApJ...341..175B}
}

@ARTICLE{2020MNRAS.498L..46V,
       author = {{Veske}, Do{\u{g}}a and {M{\'a}rka}, Zsuzsa and {Sullivan}, Andrew G. and {Bartos}, Imre and {Rainer Corley}, K. and {Samsing}, Johan and {M{\'a}rka}, Szabolcs},
        title = "{Have hierarchical three-body mergers been detected by LIGO/Virgo?}",
      journal = {\mnras},
         year = 2020,
        month = oct,
       volume = {498},
       number = {1},
        pages = {L46-L52},
          doi = {10.1093/mnrasl/slaa123},
archivePrefix = {arXiv},
       eprint = {2002.12346},
 primaryClass = {astro-ph.HE},
       adsurl = {https://ui.adsabs.harvard.edu/abs/2020MNRAS.498L..46V}
}

@ARTICLE{2025PhRvD.111j3016L,
       author = {{Li}, Guo-Peng and {Fan}, Xi-Long},
        title = "{Multimessenger hierarchical triple merger gravitational-wave event pair GW190514-GW190521 inside AGN J124942.3+344929}",
      journal = {\prd},
         year = 2025,
        month = may,
       volume = {111},
       number = {10},
          eid = {103016},
        pages = {103016},
          doi = {10.1103/PhysRevD.111.103016},
archivePrefix = {arXiv},
       eprint = {2503.16864},
 primaryClass = {astro-ph.HE},
       adsurl = {https://ui.adsabs.harvard.edu/abs/2025PhRvD.111j3016L}
}

@ARTICLE{2023arXiv230709097G,
       author = {{Gayathri}, V. and {Bartos}, I. and {Rosswog}, S. and {Miller}, M.~C. and {Veske}, D. and {Lu}, W. and {Marka}, S.},
        title = "{Do gravitational wave observations in the lower mass gap favor a hierarchical triple origin?}",
      journal = {arXiv e-prints},
         year = 2023,
        month = jul,
          eid = {arXiv:2307.09097},
        pages = {arXiv:2307.09097},
          doi = {10.48550/arXiv.2307.09097},
archivePrefix = {arXiv},
       eprint = {2307.09097},
 primaryClass = {astro-ph.HE},
       adsurl = {https://ui.adsabs.harvard.edu/abs/2023arXiv230709097G}
}

@ARTICLE{2021ApJ...907L..48V,
       author = {{Veske}, Do{\u{g}}a and {Sullivan}, Andrew G. and {M{\'a}rka}, Zsuzsa and {Bartos}, Imre and {Corley}, K. Rainer and {Samsing}, Johan and {Buscicchio}, Riccardo and {M{\'a}rka}, Szabolcs},
        title = "{Search for Black Hole Merger Families}",
      journal = {\apjl},
         year = 2021,
        month = feb,
       volume = {907},
       number = {2},
          eid = {L48},
        pages = {L48},
          doi = {10.3847/2041-8213/abd721},
archivePrefix = {arXiv},
       eprint = {2011.06591},
 primaryClass = {astro-ph.HE},
       adsurl = {https://ui.adsabs.harvard.edu/abs/2021ApJ...907L..48V}
}

@ARTICLE{2022MNRAS.511.1362T,
       author = {{Trani}, Alessandro A. and {Rastello}, Sara and {Di Carlo}, Ugo N. and {Santoliquido}, Filippo and {Tanikawa}, Ataru and {Mapelli}, Michela},
        title = "{Compact object mergers in hierarchical triples from low-mass young star clusters}",
      journal = {\mnras},
         year = 2022,
        month = feb,
       volume = {511},
       number = {1},
        pages = {1362-1372},
          doi = {10.1093/mnras/stac122},
archivePrefix = {arXiv},
       eprint = {2111.06388},
 primaryClass = {astro-ph.HE},
       adsurl = {https://ui.adsabs.harvard.edu/abs/2022MNRAS.511.1362T}
}

@ARTICLE{2026MNRAS.545f1938D,
       author = {{Dorozsmai}, Andris and {Romero-Shaw}, Isobel M. and {Vijaykumar}, Aditya and {Toonen}, Silvia and {Antonini}, Fabio and {Kremer}, Kyle and {Zevin}, Michael and {Grishin}, Evgeni},
        title = "{Hierarchical triples versus globular clusters: binary black hole merger eccentricity distributions compete and evolve with redshift}",
      journal = {\mnras},
         year = 2026,
        month = jan,
       volume = {545},
       number = {2},
          eid = {staf1938},
        pages = {staf1938},
          doi = {10.1093/mnras/staf1938},
archivePrefix = {arXiv},
       eprint = {2507.23212},
 primaryClass = {astro-ph.GA},
       adsurl = {https://ui.adsabs.harvard.edu/abs/2026MNRAS.545f1938D}
}

@ARTICLE{2016ApJ...831..187A,
       author = {{Antonini}, Fabio and {Rasio}, Frederic A.},
        title = "{Merging Black Hole Binaries in Galactic Nuclei: Implications for Advanced-LIGO Detections}",
      journal = {\apj},
         year = 2016,
        month = nov,
       volume = {831},
       number = {2},
          eid = {187},
        pages = {187},
          doi = {10.3847/0004-637X/831/2/187},
archivePrefix = {arXiv},
       eprint = {1606.04889},
 primaryClass = {astro-ph.HE},
       adsurl = {https://ui.adsabs.harvard.edu/abs/2016ApJ...831..187A}
}

@ARTICLE{2025A&A...699A.272V,
       author = {{Vigna-G{\'o}mez}, A. and {Grishin}, E. and {Stegmann}, J. and {Olejak}, A. and {Popa}, S.~A. and {Liu}, B. and {Rajamuthukumar}, A.~S. and {van Son}, L.~A.~C. and {Bobrick}, A. and {Dorozsmai}, A.},
        title = "{Prompt stellar and binary black hole mergers in tight triples: Insights from chemically homogeneous evolution}",
      journal = {\aap},
         year = 2025,
        month = jul,
       volume = {699},
          eid = {A272},
        pages = {A272},
          doi = {10.1051/0004-6361/202554680},
archivePrefix = {arXiv},
       eprint = {2503.17006},
 primaryClass = {astro-ph.SR},
       adsurl = {https://ui.adsabs.harvard.edu/abs/2025A&A...699A.272V}
}

@ARTICLE{2026arXiv260625304R,
       author = {{Roy}, Labani and {Nitz}, Alexander H.},
        title = "{Constraints on Line-of-Sight Acceleration from O1-O4}",
      journal = {arXiv e-prints},
         year = 2026,
        month = jun,
          eid = {arXiv:2606.25304},
        pages = {arXiv:2606.25304},
          doi = {10.48550/arXiv.2606.25304},
archivePrefix = {arXiv},
       eprint = {2606.25304},
 primaryClass = {astro-ph.HE},
       adsurl = {https://ui.adsabs.harvard.edu/abs/2026arXiv260625304R}
}

@ARTICLE{2020ApJ...900L..13A,
       author = {{Abbott}, R. and {Abbott}, T.~D. and {Abraham}, S. and {Acernese}, F. and {Ackley}, K. and {Adams}, C. and {Adhikari}, R.~X. and {Adya}, V.~B. and {Affeldt}, C. and {Agathos}, M. and {Agatsuma}, K. and {Aggarwal}, N. and {Aguiar}, O.~D. and {Aich}, A. and {Aiello}, L. and {Ain}, A. and {Ajith}, P. and {Akcay}, S. and {Allen}, G. and {Allocca}, A. and {Altin}, P.~A. and {Amato}, A. and {Anand}, S. and {Ananyeva}, A. and {Anderson}, S.~B. and {Anderson}, W.~G. and {Angelova}, S.~V. and {Ansoldi}, S. and {Antier}, S. and {Appert}, S. and {Arai}, K. and {Araya}, M.~C. and {Areeda}, J.~S. and {Ar{\`e}ne}, M. and {Arnaud}, N. and {Aronson}, S.~M. and {Arun}, K.~G. and {Asali}, Y. and {Ascenzi}, S. and {Ashton}, G. and {Aston}, S.~M. and {Astone}, P. and {Aubin}, F. and {Aufmuth}, P. and {AultONeal}, K. and {Austin}, C. and {Avendano}, V. and {Babak}, S. and {Bacon}, P. and {Badaracco}, F. and {Bader}, M.~K.~M. and {Bae}, S. and {Baer}, A.~M. and {Baird}, J. and {Baldaccini}, F. and {Ballardin}, G. and {Ballmer}, S.~W. and {Bals}, A. and {Balsamo}, A. and {Baltus}, G. and {Banagiri}, S. and {Bankar}, D. and {Bankar}, R.~S. and {Barayoga}, J.~C. and {Barbieri}, C. and {Barish}, B.~C. and {Barker}, D. and {Barkett}, K. and {Barneo}, P. and {Barone}, F. and {Barr}, B. and {Barsotti}, L. and {Barsuglia}, M. and {Barta}, D. and {Bartlett}, J. and {Bartos}, I. and {Bassiri}, R. and {Basti}, A. and {Bawaj}, M. and {Bayley}, J.~C. and {Bazzan}, M. and {B{\'e}csy}, B. and {Bejger}, M. and {Belahcene}, I. and {Bell}, A.~S. and {Beniwal}, D. and {Benjamin}, M.~G. and {Bentley}, J.~D. and {Bergamin}, F. and {Berger}, B.~K. and {Bergmann}, G. and {Bernuzzi}, S. and {Berry}, C.~P.~L. and {Bersanetti}, D. and {Bertolini}, A. and {Betzwieser}, J. and {Bhandare}, R. and {Bhandari}, A.~V. and {Bidler}, J. and {Biggs}, E. and {Bilenko}, I.~A. and {Billingsley}, G. and {Birney}, R. and {Birnholtz}, O. and {Biscans}, S. and {Bischi}, M. and {Biscoveanu}, S. and {Bisht}, A. and {Bissenbayeva}, G. and {Bitossi}, M. and {Bizouard}, M.~A. and {Blackburn}, J.~K. and {Blackman}, J. and {Blair}, C.~D. and {Blair}, D.~G. and {Blair}, R.~M. and {Bobba}, F. and {Bode}, N. and {Boer}, M. and {Boetzel}, Y. and {Bogaert}, G. and {Bondu}, F. and {Bonilla}, E. and {Bonnand}, R. and {Booker}, P. and {Boom}, B.~A. and {Bork}, R. and {Boschi}, V. and {Bose}, S. and {Bossilkov}, V. and {Bosveld}, J. and {Bouffanais}, Y. and {Bozzi}, A. and {Bradaschia}, C. and {Brady}, P.~R. and {Bramley}, A. and {Branchesi}, M. and {Brau}, J.~E. and {Breschi}, M. and {Briant}, T. and {Briggs}, J.~H. and {Brighenti}, F. and {Brillet}, A. and {Brinkmann}, M. and {Brockill}, P. and {Brooks}, A.~F. and {Brooks}, J. and {Brown}, D.~D. and {Brunett}, S. and {Bruno}, G. and {Bruntz}, R. and {Buikema}, A. and {Bulik}, T. and {Bulten}, H.~J. and {Buonanno}, A. and {Buscicchio}, R. and {Buskulic}, D. and {Byer}, R.~L. and {Cabero}, M. and {Cadonati}, L. and {Cagnoli}, G. and {Cahillane}, C. and {Bustillo}, J. Calder{\'o}n and {Callaghan}, J.~D. and {Callister}, T.~A. and {Calloni}, E. and {Camp}, J.~B. and {Canepa}, M. and {Cannon}, K.~C. and {Cao}, H. and {Cao}, J. and {Carapella}, G. and {Carbognani}, F. and {Caride}, S. and {Carney}, M.~F. and {Carullo}, G. and {Diaz}, J. Casanueva and {Casentini}, C. and {Casta{\~n}eda}, J. and {Caudill}, S. and {Cavagli{\`a}}, M. and {Cavalier}, F. and {Cavalieri}, R. and {Cella}, G. and {Cerd{\'a}-Dur{\'a}n}, P. and {Cesarini}, E. and {Chaibi}, O. and {Chakravarti}, K. and {Chan}, C. and {Chan}, M. and {Chao}, S. and {Charlton}, P. and {Chase}, E.~A. and {Chassande-Mottin}, E. and {Chatterjee}, D. and {Chaturvedi}, M. and {Chatziioannou}, K. and {Chen}, H.~Y. and {Chen}, X. and {Chen}, Y.},
        title = "{Properties and Astrophysical Implications of the 150 M$_{☉}$ Binary Black Hole Merger GW190521}",
      journal = {\apjl},
         year = 2020,
        month = sep,
       volume = {900},
       number = {1},
          eid = {L13},
        pages = {L13},
          doi = {10.3847/2041-8213/aba493},
archivePrefix = {arXiv},
       eprint = {2009.01190},
 primaryClass = {astro-ph.HE},
       adsurl = {https://ui.adsabs.harvard.edu/abs/2020ApJ...900L..13A}
}

@ARTICLE{2024ApJ...970L..34A,
       author = {{Abac}, A.~G. and {Abbott}, R. and {Abouelfettouh}, I. and {Acernese}, F. and {Ackley}, K. and {Adhicary}, S. and {Adhikari}, N. and {Adhikari}, R.~X. and {Adkins}, V.~K. and {Agarwal}, D. and {Agathos}, M. and {Abchouyeh}, M. Aghaei and {Aguiar}, O.~D. and {Aguilar}, I. and {Aiello}, L. and {Ain}, A. and {Ajith}, P. and {Ak{\c{c}}ay}, S. and {Akutsu}, T. and {Albanesi}, S. and {Alfaidi}, R.~A. and {Al-Jodah}, A. and {All{\'e}n{\'e}}, C. and {Allocca}, A. and {Al-Shammari}, S. and {Altin}, P.~A. and {Alvarez-Lopez}, S. and {Amato}, A. and {Amez-Droz}, L. and {Amorosi}, A. and {Amra}, C. and {Ananyeva}, A. and {Anderson}, S.~B. and {Anderson}, W.~G. and {Andia}, M. and {Ando}, M. and {Andrade}, T. and {Andres}, N. and {Andr{\'e}s-Carcasona}, M. and {Andri{\'c}}, T. and {Anglin}, J. and {Ansoldi}, S. and {Antelis}, J.~M. and {Antier}, S. and {Aoumi}, M. and {Appavuravther}, E.~Z. and {Appert}, S. and {Apple}, S.~K. and {Arai}, K. and {Araya}, A. and {Araya}, M.~C. and {Areeda}, J.~S. and {Argianas}, L. and {Aritomi}, N. and {Armato}, F. and {Arnaud}, N. and {Arogeti}, M. and {Aronson}, S.~M. and {Arun}, K.~G. and {Ashton}, G. and {Aso}, Y. and {Assiduo}, M. and {Assis de Souza Melo}, S. and {Aston}, S.~M. and {Astone}, P. and {Attadio}, F. and {Aubin}, F. and {Aultoneal}, K. and {Avallone}, G. and {Azrad}, D. and {Babak}, S. and {Badaracco}, F. and {Badger}, C. and {Bae}, S. and {Bagnasco}, S. and {Bagui}, E. and {Baier}, J.~G. and {Baiotti}, L. and {Bajpai}, R. and {Baka}, T. and {Ball}, M. and {Ballardin}, G. and {Ballmer}, S.~W. and {Banagiri}, S. and {Banerjee}, B. and {Bankar}, D. and {Baral}, P. and {Barayoga}, J.~C. and {Barish}, B.~C. and {Barker}, D. and {Barneo}, P. and {Barone}, F. and {Barr}, B. and {Barsotti}, L. and {Barsuglia}, M. and {Barta}, D. and {Bartoletti}, A.~M. and {Barton}, M.~A. and {Bartos}, I. and {Basak}, S. and {Basalaev}, A. and {Bassiri}, R. and {Basti}, A. and {Bates}, D.~E. and {Bawaj}, M. and {Baxi}, P. and {Bayley}, J.~C. and {Baylor}, A.~C. and {Baynard}, II, P.~A. and {Bazzan}, M. and {Bedakihale}, V.~M. and {Beirnaert}, F. and {Bejger}, M. and {Belardinelli}, D. and {Bell}, A.~S. and {Benedetto}, V. and {Benoit}, W. and {Bentara}, I. and {Bentley}, J.~D. and {Yaala}, M. Ben and {Bera}, S. and {Berbel}, M. and {Bergamin}, F. and {Berger}, B.~K. and {Bernuzzi}, S. and {Beroiz}, M. and {Berry}, C.~P.~L. and {Bersanetti}, D. and {Bertolini}, A. and {Betzwieser}, J. and {Beveridge}, D. and {Bevins}, N. and {Bhandare}, R. and {Bhardwaj}, U. and {Bhatt}, R. and {Bhattacharjee}, D. and {Bhaumik}, S. and {Bhowmick}, S. and {Bianchi}, A. and {Bilenko}, I.~A. and {Billingsley}, G. and {Binetti}, A. and {Bini}, S. and {Birnholtz}, O. and {Biscoveanu}, S. and {Bisht}, A. and {Bitossi}, M. and {Bizouard}, M.-A. and {Blackburn}, J.~K. and {Blagg}, L.~A. and {Blair}, C.~D. and {Blair}, D.~G. and {Bobba}, F. and {Bode}, N. and {Boileau}, G. and {Boldrini}, M. and {Bolingbroke}, G.~N. and {Bolliand}, A. and {Bonavena}, L.~D. and {Bondarescu}, R. and {Bondu}, F. and {Bonilla}, E. and {Bonilla}, M.~S. and {Bonino}, A. and {Bonnand}, R. and {Booker}, P. and {Borchers}, A. and {Boschi}, V. and {Bose}, S. and {Bossilkov}, V. and {Boudart}, V. and {Boudon}, A. and {Bozzi}, A. and {Bradaschia}, C. and {Brady}, P.~R. and {Braglia}, M. and {Branch}, A. and {Branchesi}, M. and {Brandt}, J. and {Braun}, I. and {Breschi}, M. and {Briant}, T. and {Brillet}, A. and {Brinkmann}, M. and {Brockill}, P. and {Brockmueller}, E. and {Brooks}, A.~F. and {Brown}, B.~C. and {Brown}, D.~D. and {Brozzetti}, M.~L. and {Brunett}, S. and {Bruno}, G. and {Bruntz}, R. and {Bryant}, J. and {Bucci}, F. and {Buchanan}, J. and {Bulashenko}, O. and {Bulik}, T. and {Bulten}, H.~J. and {Buonanno}, A.},
        title = "{Observation of Gravitational Waves from the Coalescence of a 2.5─4.5 M $_{{\ensuremath{\odot}}}$ Compact Object and a Neutron Star}",
      journal = {\apjl},
         year = 2024,
        month = aug,
       volume = {970},
       number = {2},
          eid = {L34},
        pages = {L34},
          doi = {10.3847/2041-8213/ad5beb},
archivePrefix = {arXiv},
       eprint = {2404.04248},
 primaryClass = {astro-ph.HE},
       adsurl = {https://ui.adsabs.harvard.edu/abs/2024ApJ...970L..34A}
}

@ARTICLE{2020ApJ...896L..44A,
       author = {{Abbott}, R. and {Abbott}, T.~D. and {Abraham}, S. and {Acernese}, F. and {Ackley}, K. and {Adams}, C. and {Adhikari}, R.~X. and {Adya}, V.~B. and {Affeldt}, C. and {Agathos}, M. and {Agatsuma}, K. and {Aggarwal}, N. and {Aguiar}, O.~D. and {Aich}, A. and {Aiello}, L. and {Ain}, A. and {Ajith}, P. and {Akcay}, S. and {Allen}, G. and {Allocca}, A. and {Altin}, P.~A. and {Amato}, A. and {Anand}, S. and {Ananyeva}, A. and {Anderson}, S.~B. and {Anderson}, W.~G. and {Angelova}, S.~V. and {Ansoldi}, S. and {Antier}, S. and {Appert}, S. and {Arai}, K. and {Araya}, M.~C. and {Areeda}, J.~S. and {Ar{\`e}ne}, M. and {Arnaud}, N. and {Aronson}, S.~M. and {Arun}, K.~G. and {Asali}, Y. and {Ascenzi}, S. and {Ashton}, G. and {Aston}, S.~M. and {Astone}, P. and {Aubin}, F. and {Aufmuth}, P. and {AultONeal}, K. and {Austin}, C. and {Avendano}, V. and {Babak}, S. and {Bacon}, P. and {Badaracco}, F. and {Bader}, M.~K.~M. and {Bae}, S. and {Baer}, A.~M. and {Baird}, J. and {Baldaccini}, F. and {Ballardin}, G. and {Ballmer}, S.~W. and {Bals}, A. and {Balsamo}, A. and {Baltus}, G. and {Banagiri}, S. and {Bankar}, D. and {Bankar}, R.~S. and {Barayoga}, J.~C. and {Barbieri}, C. and {Barish}, B.~C. and {Barker}, D. and {Barkett}, K. and {Barneo}, P. and {Barone}, F. and {Barr}, B. and {Barsotti}, L. and {Barsuglia}, M. and {Barta}, D. and {Bartlett}, J. and {Bartos}, I. and {Bassiri}, R. and {Basti}, A. and {Bawaj}, M. and {Bayley}, J.~C. and {Bazzan}, M. and {B{\'e}csy}, B. and {Bejger}, M. and {Belahcene}, I. and {Bell}, A.~S. and {Beniwal}, D. and {Benjamin}, M.~G. and {Benkel}, R. and {Bentley}, J.~D. and {Bergamin}, F. and {Berger}, B.~K. and {Bergmann}, G. and {Bernuzzi}, S. and {Berry}, C.~P.~L. and {Bersanetti}, D. and {Bertolini}, A. and {Betzwieser}, J. and {Bhandare}, R. and {Bhandari}, A.~V. and {Bidler}, J. and {Biggs}, E. and {Bilenko}, I.~A. and {Billingsley}, G. and {Birney}, R. and {Birnholtz}, O. and {Biscans}, S. and {Bischi}, M. and {Biscoveanu}, S. and {Bisht}, A. and {Bissenbayeva}, G. and {Bitossi}, M. and {Bizouard}, M.~A. and {Blackburn}, J.~K. and {Blackman}, J. and {Blair}, C.~D. and {Blair}, D.~G. and {Blair}, R.~M. and {Bobba}, F. and {Bode}, N. and {Boer}, M. and {Boetzel}, Y. and {Bogaert}, G. and {Bondu}, F. and {Bonilla}, E. and {Bonnand}, R. and {Booker}, P. and {Boom}, B.~A. and {Bork}, R. and {Boschi}, V. and {Bose}, S. and {Bossilkov}, V. and {Bosveld}, J. and {Bouffanais}, Y. and {Bozzi}, A. and {Bradaschia}, C. and {Brady}, P.~R. and {Bramley}, A. and {Branchesi}, M. and {Brau}, J.~E. and {Breschi}, M. and {Briant}, T. and {Briggs}, J.~H. and {Brighenti}, F. and {Brillet}, A. and {Brinkmann}, M. and {Brito}, R. and {Brockill}, P. and {Brooks}, A.~F. and {Brooks}, J. and {Brown}, D.~D. and {Brunett}, S. and {Bruno}, G. and {Bruntz}, R. and {Buikema}, A. and {Bulik}, T. and {Bulten}, H.~J. and {Buonanno}, A. and {Buskulic}, D. and {Byer}, R.~L. and {Cabero}, M. and {Cadonati}, L. and {Cagnoli}, G. and {Cahillane}, C. and {Bustillo}, J. Calder{\'o}n and {Callaghan}, J.~D. and {Callister}, T.~A. and {Calloni}, E. and {Camp}, J.~B. and {Canepa}, M. and {Cannon}, K.~C. and {Cao}, H. and {Cao}, J. and {Carapella}, G. and {Carbognani}, F. and {Caride}, S. and {Carney}, M.~F. and {Carullo}, G. and {Diaz}, J. Casanueva and {Casentini}, C. and {Casta{\~n}eda}, J. and {Caudill}, S. and {Cavagli{\`a}}, M. and {Cavalier}, F. and {Cavalieri}, R. and {Cella}, G. and {Cerd{\'a}-Dur{\'a}n}, P. and {Cesarini}, E. and {Chaibi}, O. and {Chakravarti}, K. and {Chan}, C. and {Chan}, M. and {Chao}, S. and {Charlton}, P. and {Chase}, E.~A. and {Chassande-Mottin}, E. and {Chatterjee}, D. and {Chaturvedi}, M. and {Chatziioannou}, K. and {Chen}, H.~Y. and {Chen}, X.},
        title = "{GW190814: Gravitational Waves from the Coalescence of a 23 Solar Mass Black Hole with a 2.6 Solar Mass Compact Object}",
      journal = {\apjl},
         year = 2020,
        month = jun,
       volume = {896},
       number = {2},
          eid = {L44},
        pages = {L44},
          doi = {10.3847/2041-8213/ab960f},
archivePrefix = {arXiv},
       eprint = {2006.12611},
 primaryClass = {astro-ph.HE},
       adsurl = {https://ui.adsabs.harvard.edu/abs/2020ApJ...896L..44A}
}

@ARTICLE{2024ApJ...970..191A,
       author = {{Abbott}, R. and {Abe}, H. and {Acernese}, F. and {Ackley}, K. and {Adhicary}, S. and {Adhikari}, N. and {Adhikari}, R.~X. and {Adkins}, V.~K. and {Adya}, V.~B. and {Affeldt}, C. and {Agarwal}, D. and {Agathos}, M. and {Aguiar}, O.~D. and {Aiello}, L. and {Ain}, A. and {Ajith}, P. and {Akutsu}, T. and {Albanesi}, S. and {Alfaidi}, R.~A. and {All{\'e}n{\'e}}, C. and {Allocca}, A. and {Altin}, P.~A. and {Amato}, A. and {Anand}, S. and {Ananyeva}, A. and {Anderson}, S.~B. and {Anderson}, W.~G. and {Ando}, M. and {Andrade}, T. and {Andres}, N. and {Andr{\'e}s-Carcasona}, M. and {Andri{\'c}}, T. and {Ansoldi}, S. and {Antelis}, J.~M. and {Antier}, S. and {Apostolatos}, T. and {Appavuravther}, E.~Z. and {Appert}, S. and {Apple}, S.~K. and {Arai}, K. and {Araya}, A. and {Araya}, M.~C. and {Areeda}, J.~S. and {Ar{\`e}ne}, M. and {Aritomi}, N. and {Arnaud}, N. and {Arogeti}, M. and {Aronson}, S.~M. and {Asada}, H. and {Ashton}, G. and {Aso}, Y. and {Assiduo}, M. and {Assis de Souza Melo}, S. and {Aston}, S.~M. and {Astone}, P. and {Aubin}, F. and {Aultoneal}, K. and {Babak}, S. and {Badaracco}, F. and {Badger}, C. and {Bae}, S. and {Bae}, Y. and {Bagnasco}, S. and {Bai}, Y. and {Baier}, J.~G. and {Baird}, J. and {Bajpai}, R. and {Baka}, T. and {Ball}, M. and {Ballardin}, G. and {Ballmer}, S.~W. and {Baltus}, G. and {Banagiri}, S. and {Banerjee}, B. and {Bankar}, D. and {Barayoga}, J.~C. and {Barish}, B.~C. and {Barker}, D. and {Barneo}, P. and {Barone}, F. and {Barr}, B. and {Barsotti}, L. and {Barsuglia}, M. and {Barta}, D. and {Bartlett}, J. and {Barton}, M.~A. and {Bartos}, I. and {Basak}, S. and {Bassiri}, R. and {Basti}, A. and {Bawaj}, M. and {Bayley}, J.~C. and {Bazzan}, M. and {B{\'e}csy}, B. and {Bedakihale}, V.~M. and {Beirnaert}, F. and {Bejger}, M. and {Belahcene}, I. and {Bell}, A.~S. and {Benedetto}, V. and {Beniwal}, D. and {Benoit}, W. and {Bentley}, J.~D. and {Benyaala}, M. and {Bera}, S. and {Berbel}, M. and {Bergamin}, F. and {Berger}, B.~K. and {Bernuzzi}, S. and {Beroiz}, M. and {Berry}, C.~P.~L. and {Bersanetti}, D. and {Bertolini}, A. and {Betzwieser}, J. and {Beveridge}, D. and {Bhandare}, R. and {Bhandari}, A.~V. and {Bhardwaj}, U. and {Bhatt}, R. and {Bhattacharjee}, D. and {Bhaumik}, S. and {Bianchi}, A. and {Bilenko}, I.~A. and {Bilicki}, M. and {Billingsley}, G. and {Bini}, S. and {Birnholtz}, O. and {Biscans}, S. and {Bischi}, M. and {Biscoveanu}, S. and {Bisht}, A. and {Biswas}, B. and {Bitossi}, M. and {Bizouard}, M.-A. and {Blackburn}, J.~K. and {Blair}, C.~D. and {Blair}, D.~G. and {Blair}, R.~M. and {Bobba}, F. and {Bode}, N. and {Bo{\"e}r}, M. and {Bogaert}, G. and {Boldrini}, M. and {Bolingbroke}, G.~N. and {Bonavena}, L.~D. and {Bondarescu}, R. and {Bondu}, F. and {Bonilla}, E. and {Bonnand}, R. and {Booker}, P. and {Bork}, R. and {Boschi}, V. and {Bose}, N. and {Bose}, S. and {Bossilkov}, V. and {Boudart}, V. and {Bouffanais}, Y. and {Bozzi}, A. and {Bradaschia}, C. and {Brady}, P.~R. and {Bramley}, A. and {Branch}, A. and {Branchesi}, M. and {Brau}, J.~E. and {Breschi}, M. and {Briant}, T. and {Briggs}, J.~H. and {Brillet}, A. and {Brinkmann}, M. and {Brockill}, P. and {Brooks}, A.~F. and {Brooks}, J. and {Brown}, D.~D. and {Brunett}, S. and {Bruno}, G. and {Bruntz}, R. and {Bryant}, J. and {Bucci}, F. and {Buchanan}, J. and {Bulik}, T. and {Bulten}, H.~J. and {Buonanno}, A. and {Burtnyk}, K. and {Buscicchio}, R. and {Buskulic}, D. and {Buy}, C. and {Byer}, R.~L. and {Cabourn Davies}, G.~S. and {Cabras}, G. and {Cabrita}, R. and {Cadonati}, L. and {Cagnoli}, G. and {Cahillane}, C. and {Calder{\'o}n Bustillo}, J. and {Callaghan}, J.~D. and {Callister}, T.~A. and {Calloni}, E. and {Camp}, J.~B. and {Canepa}, M. and {Caneva}, G.},
        title = "{Search for Gravitational-lensing Signatures in the Full Third Observing Run of the LIGO─Virgo Network}",
      journal = {\apj},
         year = 2024,
        month = aug,
       volume = {970},
       number = {2},
          eid = {191},
        pages = {191},
          doi = {10.3847/1538-4357/ad3e83},
archivePrefix = {arXiv},
       eprint = {2304.08393},
 primaryClass = {gr-qc},
       adsurl = {https://ui.adsabs.harvard.edu/abs/2024ApJ...970..191A}
}

@ARTICLE{2021ApJ...915L..35K,
       author = {{Kimball}, Chase and {Talbot}, Colm and {Berry}, Christopher P.~L. and {Zevin}, Michael and {Thrane}, Eric and {Kalogera}, Vicky and {Buscicchio}, Riccardo and {Carney}, Matthew and {Dent}, Thomas and {Middleton}, Hannah and {Payne}, Ethan and {Veitch}, John and {Williams}, Daniel},
        title = "{Evidence for Hierarchical Black Hole Mergers in the Second LIGO-Virgo Gravitational Wave Catalog}",
      journal = {\apjl},
         year = 2021,
        month = jul,
       volume = {915},
       number = {2},
          eid = {L35},
        pages = {L35},
          doi = {10.3847/2041-8213/ac0aef},
archivePrefix = {arXiv},
       eprint = {2011.05332},
 primaryClass = {astro-ph.HE},
       adsurl = {https://ui.adsabs.harvard.edu/abs/2021ApJ...915L..35K}
}

@ARTICLE{2020A&A...635A..97B,
       author = {{Bavera}, Simone S. and {Fragos}, Tassos and {Qin}, Ying and {Zapartas}, Emmanouil and {Neijssel}, Coenraad J. and {Mandel}, Ilya and {Batta}, Aldo and {Gaebel}, Sebastian M. and {Kimball}, Chase and {Stevenson}, Simon},
        title = "{The origin of spin in binary black holes. Predicting the distributions of the main observables of Advanced LIGO}",
      journal = {\aap},
         year = 2020,
        month = mar,
       volume = {635},
          eid = {A97},
        pages = {A97},
          doi = {10.1051/0004-6361/201936204},
archivePrefix = {arXiv},
       eprint = {1906.12257},
 primaryClass = {astro-ph.HE},
       adsurl = {https://ui.adsabs.harvard.edu/abs/2020A&A...635A..97B}
}

@ARTICLE{2017CQGra..34cLT01V,
       author = {{Vitale}, Salvatore and {Lynch}, Ryan and {Sturani}, Riccardo and {Graff}, Philip},
        title = "{Use of gravitational waves to probe the formation channels of compact binaries}",
      journal = {Classical and Quantum Gravity},
         year = 2017,
        month = feb,
       volume = {34},
       number = {3},
          eid = {03LT01},
        pages = {03LT01},
          doi = {10.1088/1361-6382/aa552e},
archivePrefix = {arXiv},
       eprint = {1503.04307},
 primaryClass = {gr-qc},
       adsurl = {https://ui.adsabs.harvard.edu/abs/2017CQGra..34cLT01V}
}

@ARTICLE{2026arXiv260320430A,
       author = {{Arca Sedda}, Manuel and {Paiella}, Lavinia and {Ugolini}, Cristiano and {Santoliquido}, Filippo and {Mestichelli}, Benedetta and {Usai}, Ilaria and {Simonato}, Filippo and {Branchesi}, Marica},
        title = "{Isolated or Dynamical? Tracing Black Hole Binary Formation through the Population of Gravitational-Wave Sources}",
      journal = {arXiv e-prints},
         year = 2026,
        month = mar,
          eid = {arXiv:2603.20430},
        pages = {arXiv:2603.20430},
          doi = {10.48550/arXiv.2603.20430},
archivePrefix = {arXiv},
       eprint = {2603.20430},
 primaryClass = {astro-ph.GA},
       adsurl = {https://ui.adsabs.harvard.edu/abs/2026arXiv260320430A}
}

@ARTICLE{2025ApJ...991..131Z,
       author = {{Zwick}, Lorenz and {Tak{\'a}tsy}, J{\'a}nos and {Saini}, Pankaj and {Hendriks}, Kai and {Samsing}, Johan and {Tiede}, Christopher and {Rowan}, Connar and {Trani}, Alessandro A.},
        title = "{Environmental Effects in Stellar Mass Gravitational-wave Sources. I. Expected Fraction of Signals with Significant Dephasing in the Dynamical and Active Galactic Nucleus Channels}",
      journal = {\apj},
         year = 2025,
        month = oct,
       volume = {991},
       number = {2},
          eid = {131},
        pages = {131},
          doi = {10.3847/1538-4357/adf6b8},
archivePrefix = {arXiv},
       eprint = {2503.24084},
 primaryClass = {astro-ph.HE},
       adsurl = {https://ui.adsabs.harvard.edu/abs/2025ApJ...991..131Z}
}

@ARTICLE{2020PhRvD.102l3016A,
       author = {{Antonini}, Fabio and {Gieles}, Mark},
        title = "{Merger rate of black hole binaries from globular clusters: Theoretical error bars and comparison to gravitational wave data from GWTC-2}",
      journal = {\prd},
         year = 2020,
        month = dec,
       volume = {102},
       number = {12},
          eid = {123016},
        pages = {123016},
          doi = {10.1103/PhysRevD.102.123016},
archivePrefix = {arXiv},
       eprint = {2009.01861},
 primaryClass = {astro-ph.HE},
       adsurl = {https://ui.adsabs.harvard.edu/abs/2020PhRvD.102l3016A}
}

@ARTICLE{2021MNRAS.505..339M,
       author = {{Mapelli}, Michela and {Dall'Amico}, Marco and {Bouffanais}, Yann and {Giacobbo}, Nicola and {Arca Sedda}, Manuel and {Artale}, M. Celeste and {Ballone}, Alessandro and {Di Carlo}, Ugo N. and {Iorio}, Giuliano and {Santoliquido}, Filippo and {Torniamenti}, Stefano},
        title = "{Hierarchical black hole mergers in young, globular and nuclear star clusters: the effect of metallicity, spin and cluster properties}",
      journal = {\mnras},
         year = 2021,
        month = jul,
       volume = {505},
       number = {1},
        pages = {339-358},
          doi = {10.1093/mnras/stab1334},
archivePrefix = {arXiv},
       eprint = {2103.05016},
 primaryClass = {astro-ph.HE},
       adsurl = {https://ui.adsabs.harvard.edu/abs/2021MNRAS.505..339M}
}

@ARTICLE{2018arXiv180707062H,
       author = {{Haris}, K. and {Mehta}, Ajit Kumar and {Kumar}, Sumit and {Venumadhav}, Tejaswi and {Ajith}, Parameswaran},
        title = "{Identifying strongly lensed gravitational wave signals from binary black hole mergers}",
      journal = {arXiv e-prints},
         year = 2018,
        month = jul,
          eid = {arXiv:1807.07062},
        pages = {arXiv:1807.07062},
          doi = {10.48550/arXiv.1807.07062},
archivePrefix = {arXiv},
       eprint = {1807.07062},
 primaryClass = {gr-qc},
       adsurl = {https://ui.adsabs.harvard.edu/abs/2018arXiv180707062H}
}

@ARTICLE{2025PhRvD.112d4038G,
       author = {{Gamboa}, Aldo and {Buonanno}, Alessandra and {Enficiaud}, Raffi and {Khalil}, Mohammed and {Ramos-Buades}, Antoni and {Pompili}, Lorenzo and {Estell{\'e}s}, H{\'e}ctor and {Boyle}, Michael and {Kidder}, Lawrence E. and {Pfeiffer}, Harald P. and {R{\"u}ter}, Hannes R. and {Scheel}, Mark A.},
        title = "{Accurate waveforms for eccentric, aligned-spin binary black holes: The multipolar effective-one-body model SEOBNRv5EHM}",
      journal = {\prd},
         year = 2025,
        month = aug,
       volume = {112},
       number = {4},
          eid = {044038},
        pages = {044038},
          doi = {10.1103/jxrc-z298},
archivePrefix = {arXiv},
       eprint = {2412.12823},
 primaryClass = {gr-qc},
       adsurl = {https://ui.adsabs.harvard.edu/abs/2025PhRvD.112d4038G}
}

@ARTICLE{2026arXiv260527226T,
       author = {{The LIGO Scientific Collaboration} and {the Virgo Collaboration} and {the KAGRA Collaboration} and {Abac}, A.~G. and {Abe}, A. and {Abouelfettouh}, I. and {Acernese}, F. and {Ackley}, K. and {Adam}, A. and {Adhicary}, S. and {Adhikari}, D. and {Adhikari}, R.~X. and {Adkins}, V.~K. and {Afroz}, S. and {Agapito}, A. and {Agarwal}, D. and {Agathos}, M. and {Aggarwal}, N. and {Aggarwal}, S. and {Aguiar}, O.~D. and {Ahrend}, I.-L. and {Aiello}, L. and {Ain}, A. and {Ajith}, P. and {Akutsu}, T. and {Albers}, L. and {Ali}, W. and {Al-Kershi}, S. and {Allene}, C. and {Allocca}, A. and {Al-Shammari}, S. and {Alvarez}, J.~A. and {Alvarez-Lopez}, S. and {Amar}, W. and {Amarasinghe}, O. and {Amato}, A. and {Amicucci}, F. and {Amra}, C. and {Anand}, A.~B. and {Anand}, C. and {Ananyeva}, A. and {Anderson}, S.~B. and {Anderson}, W.~G. and {Andia}, M. and {Ando}, M. and {Andrade-Oliveira}, F. and {Andr{\'e}s-Carcasona}, M. and {Andrey}, J.~L. and {Andri{\'c}}, T. and {Anglin}, J. and {Anna}, J. and {Antelis}, J.~M. and {Antier}, S. and {Antonini}, F. and {Aoki}, T. and {Aoumi}, M. and {Appavuravther}, E.~Z. and {Appelt}, E.~A. and {Appert}, S. and {Apple}, S.~K. and {Arai}, K. and {Araya}, A. and {Araya}, M.~C. and {Arca Sedda}, M. and {Arciprete}, F. and {Areeda}, J.~S. and {Aritomi}, N. and {Armato}, F. and {Armstrong}, S. and {Arnaud}, N. and {Arogeti}, M. and {Aronson}, S.~M. and {Ashton}, G. and {Aso}, Y. and {Asprea}, L. and {Assiduo}, M. and {Assis de Souza Melo}, S. and {Aston}, S.~M. and {Astone}, P. and {Aswathi}, P.~S. and {Attadio}, F. and {Aubin}, F. and {AultONeal}, K. and {Avallone}, G. and {Avdeev}, N. and {Avila}, E.~A. and {Babak}, S. and {Badger}, C. and {Bae}, S. and {Bagnasco}, S. and {Baimukhametova}, S. and {Baiotti}, L. and {Baka}, T. and {Baker}, K.~A. and {Baker}, T. and {Balbi}, G. and {Baldi}, G. and {Baldicchi}, N. and {Ball}, M. and {Ballardin}, G. and {Ballelli}, M. and {Ballmer}, S.~W. and {Banagiri}, S. and {Banerjee}, B. and {Bankar}, D. and {Baptiste}, T.~M. and {Baral}, P. and {Baratti}, M. and {Barayoga}, J.~C. and {Baric}, K. and {Barish}, B.~C. and {Barker}, D. and {Barman}, N. and {Barone}, F. and {Barr}, B. and {Barrios}, M. and {Barsotti}, L. and {Barsuglia}, M. and {Barta}, D. and {Barton}, M.~A. and {Bartos}, I. and {Basalaev}, A. and {Bassiri}, R. and {Basti}, A. and {Bawaj}, M. and {Bayley}, J.~C. and {Baylor}, A.~C. and {Baynard}, II, P.~A. and {Bazzan}, M. and {Bedakihale}, V.~M. and {Beirnaert}, F. and {Bejger}, M. and {Bell}, A.~S. and {Bellani}, C. and {Bellie}, D.~S. and {Beltran-Martinez}, D. and {Benedetti}, E. and {Benoit}, W. and {Bentara}, I. and {Ben Yaala}, M. and {Bera}, S. and {Bergamin}, F. and {Berger}, B.~K. and {Beroiz}, M. and {Berry}, C.~P.~L. and {Berry}, I. and {Bersanetti}, D. and {Bertheas}, T. and {Bertolini}, A. and {Betzwieser}, J. and {Beveridge}, D. and {Bevins}, N. and {Bezerra-Sobrinho}, J. and {Bhandare}, R. and {Bhatt}, R. and {Bhattacharjee}, A. and {Bhattacharjee}, D. and {Bhattacharyya}, S. and {Bhaumik}, S. and {Biancalana}, V. and {Bianchi}, F. and {Bilenko}, I.~A. and {Bilicki}, M. and {Billingsley}, G. and {Binetti}, A. and {Bini}, S. and {Biot}, S. and {Birnholtz}, O. and {Biscoveanu}, S. and {Bisht}, A. and {Bitossi}, M. and {Bizouard}, M.-A. and {Blaber}, S. and {Blackburn}, J.~K. and {Blagg}, L.~A. and {Blair}, C.~D. and {Blair}, D.~G. and {Bloch}, M. and {Bode}, N. and {Boettner}, N. and {Bogdan}, P. and {Boileau}, G. and {Boldrini}, M. and {Bolingbroke}, G.~N. and {Bonavena}, L.~D. and {Bonhomme}, V.~A. and {Bonilla}, E. and {Bonilla}, M.~S. and {Bonino}, A. and {Bonnand}, R. and {Borchers}, A. and {Borghi}, N. and {Boschi}, V. and {Bose}, S. and {Bossilkov}, V. and {Bothra}, Y. and {Boudon}, A. and {Boybeyi}, T.~D. and {Boyle}, M. and {Bozzi}, A.},
        title = "{GWTC-5.0: Population Properties of Merging Compact Binaries}",
      journal = {arXiv e-prints},
         year = 2026,
        month = may,
          eid = {arXiv:2605.27226},
        pages = {arXiv:2605.27226},
          doi = {10.48550/arXiv.2605.27226},
archivePrefix = {arXiv},
       eprint = {2605.27226},
 primaryClass = {astro-ph.HE},
       adsurl = {https://ui.adsabs.harvard.edu/abs/2026arXiv260527226T}
}

@ARTICLE{2016ApJ...833L...1A,
       author = {{Abbott}, B.~P. and {Abbott}, R. and {Abbott}, T.~D. and {Abernathy}, M.~R. and {Acernese}, F. and {Ackley}, K. and {Adams}, C. and {Adams}, T. and {Addesso}, P. and {Adhikari}, R.~X. and {Adya}, V.~B. and {Affeldt}, C. and {Agathos}, M. and {Agatsuma}, K. and {Aggarwal}, N. and {Aguiar}, O.~D. and {Aiello}, L. and {Ain}, A. and {Ajith}, P. and {Allen}, B. and {Allocca}, A. and {Altin}, P.~A. and {Anderson}, S.~B. and {Anderson}, W.~G. and {Arai}, K. and {Araya}, M.~C. and {Arceneaux}, C.~C. and {Areeda}, J.~S. and {Arnaud}, N. and {Arun}, K.~G. and {Ascenzi}, S. and {Ashton}, G. and {Ast}, M. and {Aston}, S.~M. and {Astone}, P. and {Aufmuth}, P. and {Aulbert}, C. and {Babak}, S. and {Bacon}, P. and {Bader}, M.~K.~M. and {Baker}, P.~T. and {Baldaccini}, F. and {Ballardin}, G. and {Ballmer}, S.~W. and {Barayoga}, J.~C. and {Barclay}, S.~E. and {Barish}, B.~C. and {Barker}, D. and {Barone}, F. and {Barr}, B. and {Barsotti}, L. and {Barsuglia}, M. and {Barta}, D. and {Bartlett}, J. and {Bartos}, I. and {Bassiri}, R. and {Basti}, A. and {Batch}, J.~C. and {Baune}, C. and {Bavigadda}, V. and {Bazzan}, M. and {Behnke}, B. and {Bejger}, M. and {Bell}, A.~S. and {Bell}, C.~J. and {Berger}, B.~K. and {Bergman}, J. and {Bergmann}, G. and {Berry}, C.~P.~L. and {Bersanetti}, D. and {Bertolini}, A. and {Betzwieser}, J. and {Bhagwat}, S. and {Bhandare}, R. and {Bilenko}, I.~A. and {Billingsley}, G. and {Birch}, J. and {Birney}, R. and {Biscans}, S. and {Bisht}, A. and {Bitossi}, M. and {Biwer}, C. and {Bizouard}, M.~A. and {Blackburn}, J.~K. and {Blair}, C.~D. and {Blair}, D.~G. and {Blair}, R.~M. and {Bloemen}, S. and {Bock}, O. and {Bodiya}, T.~P. and {Boer}, M. and {Bogaert}, G. and {Bogan}, C. and {Bohe}, A. and {Bojtos}, P. and {Bond}, C. and {Bondu}, F. and {Bonnand}, R. and {Boom}, B.~A. and {Bork}, R. and {Boschi}, V. and {Bose}, S. and {Bouffanais}, Y. and {Bozzi}, A. and {Bradaschia}, C. and {Brady}, P.~R. and {Braginsky}, V.~B. and {Branchesi}, M. and {Brau}, J.~E. and {Briant}, T. and {Brillet}, A. and {Brinkmann}, M. and {Brisson}, V. and {Brockill}, P. and {Brooks}, A.~F. and {Brown}, D.~A. and {Brown}, D.~D. and {Brown}, N.~M. and {Buchanan}, C.~C. and {Buikema}, A. and {Bulik}, T. and {Bulten}, H.~J. and {Buonanno}, A. and {Buskulic}, D. and {Buy}, C. and {Byer}, R.~L. and {Cadonati}, L. and {Cagnoli}, G. and {Cahillane}, C. and {Calder{\'o}n Bustillo}, J. and {Callister}, T. and {Calloni}, E. and {Camp}, J.~B. and {Cannon}, K.~C. and {Cao}, J. and {Capano}, C.~D. and {Capocasa}, E. and {Carbognani}, F. and {Caride}, S. and {Casanueva Diaz}, J. and {Casentini}, C. and {Caudill}, S. and {Cavagli{\`a}}, M. and {Cavalier}, F. and {Cavalieri}, R. and {Cella}, G. and {Cepeda}, C.~B. and {Cerboni Baiardi}, L. and {Cerretani}, G. and {Cesarini}, E. and {Chakraborty}, R. and {Chalermsongsak}, T. and {Chamberlin}, S.~J. and {Chan}, M. and {Chao}, S. and {Charlton}, P. and {Chassande-Mottin}, E. and {Chen}, H.~Y. and {Chen}, Y. and {Cheng}, C. and {Chincarini}, A. and {Chiummo}, A. and {Cho}, H.~S. and {Cho}, M. and {Chow}, J.~H. and {Christensen}, N. and {Chu}, Q. and {Chua}, S. and {Chung}, S. and {Ciani}, G. and {Clara}, F. and {Clark}, J.~A. and {Cleva}, F. and {Coccia}, E. and {Cohadon}, P.-F. and {Colla}, A. and {Collette}, C.~G. and {Cominsky}, L. and {Constancio}, Jr., M. and {Conte}, A. and {Conti}, L. and {Cook}, D. and {Corbitt}, T.~R. and {Cornish}, N. and {Corsi}, A. and {Cortese}, S. and {Costa}, C.~A. and {Coughlin}, M.~W. and {Coughlin}, S.~B. and {Coulon}, J.-P. and {Countryman}, S.~T. and {Couvares}, P. and {Cowan}, E.~E. and {Coward}, D.~M. and {Cowart}, M.~J. and {Coyne}, D.~C. and {Coyne}, R. and {Craig}, K. and {Creighton}, J.~D.~E. and {Cripe}, J.},
        title = "{The Rate of Binary Black Hole Mergers Inferred from Advanced LIGO Observations Surrounding GW150914}",
      journal = {\apjl},
         year = 2016,
        month = dec,
       volume = {833},
       number = {1},
          eid = {L1},
        pages = {L1},
          doi = {10.3847/2041-8205/833/1/L1},
archivePrefix = {arXiv},
       eprint = {1602.03842},
 primaryClass = {astro-ph.HE},
       adsurl = {https://ui.adsabs.harvard.edu/abs/2016ApJ...833L...1A}
}

@ARTICLE{2026ApJ...999..127L,
       author = {{Li}, Yin-Jie and {Tang}, Shao-Peng and {Xue}, Ling-Qin and {Fan}, Yi-Zhong},
        title = "{GW231123: Likely a Product of Successive Mergers from {\ensuremath{\sim}}10 Stellar-mass Black Holes}",
      journal = {\apj},
         year = 2026,
        month = mar,
       volume = {999},
       number = {1},
          eid = {127},
        pages = {127},
          doi = {10.3847/1538-4357/ae4102},
archivePrefix = {arXiv},
       eprint = {2507.17551},
 primaryClass = {astro-ph.HE},
       adsurl = {https://ui.adsabs.harvard.edu/abs/2026ApJ...999..127L}
}

@ARTICLE{2024PhRvL.133e1401L,
       author = {{Li}, Yin-Jie and {Wang}, Yuan-Zhu and {Tang}, Shao-Peng and {Fan}, Yi-Zhong},
        title = "{Resolving the Stellar-Collapse and Hierarchical-Merger Origins of the Coalescing Black Holes}",
      journal = {\prl},
         year = 2024,
        month = aug,
       volume = {133},
       number = {5},
          eid = {051401},
        pages = {051401},
          doi = {10.1103/PhysRevLett.133.051401},
archivePrefix = {arXiv},
       eprint = {2303.02973},
 primaryClass = {astro-ph.HE},
       adsurl = {https://ui.adsabs.harvard.edu/abs/2024PhRvL.133e1401L}
}

@ARTICLE{2023PhRvD.108d3011E,
       author = {{Essick}, Reed},
        title = "{Semianalytic sensitivity estimates for catalogs of gravitational-wave transients}",
      journal = {\prd},
         year = 2023,
        month = aug,
       volume = {108},
       number = {4},
          eid = {043011},
        pages = {043011},
          doi = {10.1103/PhysRevD.108.043011},
archivePrefix = {arXiv},
       eprint = {2307.02765},
 primaryClass = {gr-qc},
       adsurl = {https://ui.adsabs.harvard.edu/abs/2023PhRvD.108d3011E}
}
\bibliographystyle{aasjournalv7}

\end{document}